\documentclass[twocolumn,amsmath,amssymb,floatfix,superscriptaddress,prx]{revtex4-2}
\usepackage{bm,graphicx,url,epsf,color}
\usepackage{amssymb}
\usepackage{amsthm}
\usepackage[colorlinks=true,linkcolor=blue,citecolor=blue]{hyperref}
\usepackage{comment}

\newcommand{\cm}[1]{{\color{black}{#1}}}
\newcommand{\HS}[1]{{\color{black}{#1}}}

\begin{document}

\title{Fractionalization and anyonic statistics in the integer quantum Hall collider}

\author{Tom Morel}
\affiliation{Universit\'e  de  Paris, Laboratoire  Mat\'eriaux  et  Ph\'enom\`enes  Quantiques, CNRS, 75013  Paris,  France}
\author{June-Young M. Lee}
\affiliation{Department of Physics, Korea Advanced Institute of Science and Technology, Daejeon 34141, Korea}
\author{H.-S. Sim} \email[]{hssim@kaist.ac.kr}
\affiliation{Department of Physics, Korea Advanced Institute of Science and Technology, Daejeon 34141, Korea}
\author{Christophe Mora}\email[]{christophe.mora@u-paris.fr}
\affiliation{Universit\'e  de  Paris, Laboratoire  Mat\'eriaux  et  Ph\'enom\`enes  Quantiques, CNRS, 75013  Paris,  France}
\affiliation{Dahlem Center for Complex Quantum Systems and Fachbereich Physik, Freie Universit\"at Berlin, 14195, Berlin, Germany}

\begin{abstract}
One remarkable feature of strongly correlated systems is the phenomenon of fractionalization where quasiparticles carry only a fraction of the charge or spin of the elementary constituents. Such quasiparticles often present anyonic statistics in two dimensions and lie at the heart of the fractional quantum Hall effect. We discuss the observation of fractionalization and anyonic statistics already in the integer quantum Hall effect coupled to a metallic island. A continuous fractional emitter is proposed, which sends dilute beams of non-integer charges, and its full counting statistics is obtained. The fractional charge is governed solely by the number of ballistic channels covered by the island and it is one half of the electron charge for a single ballistic channel. We further characterize the mixing of two such fractional beams through a quantum point contact beam splitter. We predict negative cross-correlations, in strong contrast with free electrons, that depend on the double exchange phase between electrons and the fractional charges emulating anyons. The result is similar to a  genuine fractional edge state as recently measured at filling $\nu = 1/3$. We revisit the physical interpretation of this experiment and point towards a direct braiding measurement rather than a deviation from fermionic antibunching.
\end{abstract}

\maketitle
\section{Introduction}

The hallmark of the fractional quantum Hall effect~\cite{tsui1982} is the existence of quasiparticles which carry a fractional charge and exhibit anyonic statistics~\cite{stern2008}. They are associated with a topological order and ground state degeneracy~\cite{wen1995}. Their exchange properties differ from fermions or bosons, possessing a statistical anyonic phase for abelian states~\cite{Halperin1984,arovas1984}. In the  non-abelian case, their braiding properties makes them building blocks for realizing fault-tolerant quantum computation~\cite{nayak2008}. Anyons reside in the bulk or are delocalized at the edge of the system where they carry an electric current and give a fractional quantized Hall conductance~\cite{halperin1982,wen1990}.
Aside from these consistent theoretical constructions, the experimental direct evidence of anyonic properties posed and still poses its own challenges~\cite{heiblum2020}. The first strong indication of their existence was obtained almost 25 years ago~\cite{DePicciotto1997,Saminadayar1997} in seminal noise measurements in the $\nu=1/3$ state. The weak tunneling between opposite edges, via a quantum point contact, involves fractional charges detected in the shot noise. The fractional charge was further confirmed in recent experiments with finite-frequency noise~\cite{bisognin2019} or photo-assisted shot noise~\cite{kapfer2019}, building an overall convincing picture.

Beyond the measurement of the fractional charge and further successes in observing quantized thermal transport~\cite{Jezouin2013b,Banerjee2017,Banerjee2018,Srivastav2019,Srivastav2021}, the experimental signature of braiding statistics of anyons remained  for a long time elusive~\cite{willett2009}, despite several proposals of interferometric~\cite{law2006,chamon1997,heiblum2020} or correlation noise~\cite{safi2001,vishveshwara2003,kim2005,campagnano2012,lee2019} probes. Two recent breakthrough experiments filled this gap and observe convincing signatures of anyonic statistics for the $\nu=1/3$ state. In the first experiment ~\cite{Bartolomei2020} following a theoretical proposal~\cite{rosenow2016}, a collider of two dilute beams of anyons propagating along the edge was realized in a GaAs/AlGaAs heterostructure. Cross-correlations after the quantum point contact, {\it i.e.} the collider, were measured to be negative in precise quantitative agreement with theory, in a way that depends on the anyonic braid phase. For comparison, cross-correlations vanish for fermions as confirmed in the same experiment. The second experiment~\cite{Nakamura2020} measured the interference fringes of a Fabry-Perot in a two-dimensional electron gas with strong screening. The path of interference encircles the bulk and thus reflects the phase accumulated by braiding one anyon on the edge with each individual bulk anyon. The abelian braid phase $2 \pi/3$, expected at filling $1/3$, was then detected upon adding one anyon to the bulk. 

In this paper, we show that anyonic statistics emerge also in the integer quantum Hall effect with a metallic island, yielding a negative cross-correlation in the same collider geometry as Ref.~\cite{Bartolomei2020}. We first devise a fractional emitter by connecting a metallic island to $N$ chiral ballistic channels and one quasi-ballistic channel using a nearby quantum point contact, and prove that it emits fractional charges $e^* = N e/(N+1)$ and $e^* = e/(N+1)$ with poissonian statistics.  The emitted quasiparticles possess a fractional charge but also emulate fractional anyon-like commutation relations. 
This emulation should yield a measurable exchange phase shift in the interference pattern of a Mach-Zehnder geometry, as shown in Ref.~\cite{lee2020} with a similar emitter.
We characterize in this work the mixing of two fractional beams originating from our emitters by computing the output noise cross-correlations. We find them to be negative in contrast with the result for non-interacting fermions and similarly to the fractional quantum Hall case~\cite{rosenow2016}. The obtained cross-correlations depend on the mutual statistical phase between electrons and emulated anyons and give a different functional form than in the case of genuine fractional edge states where only anyons are (braided) exchanged. We further elaborate on the physical picture behind these cross-correlations and find that they account directly for the double exchange between an electron and a fractional charge (anyon), whereas the (anti-)bunching effect in the collision gives only a subleading neglected term. Interestingly, we find that the same is true in the fractional quantum Hall state discussed in Ref.~\cite{rosenow2016}, such that the corresponding experiment~\cite{Bartolomei2020} should probably be interpreted as a direct consequence 
of the anyonic braid phase and not as a deviation to fermionic anti-bunching, the corresponding contribution being small.

To make contact with previous works, we note that the fractional emitter can alternatively be seen as a conductor, defined by the quasi-ballistic channel, experiencing dynamical Coulomb blockade from a resistive environment, composed of the $N$ ballistic channels and the dot~\cite{Parmentier2011,duprez2021}. The model can further be mapped onto an impurity in a Luttinger liquid~\cite{safi2004,Jezouin2013,Anthore2018} where non-integer charges are known to occur~\cite{safi1995,Pham2000,Steinberg2008}. By continuously changing the couplings between the island and the edge channels, the model for the emitter also evolves into a charge Kondo model~\cite{Matveev1991,Matveev1995,Furusaki1995b,Iftikhar2015,Iftikhar2018}, and fractional charges have been discussed in this context too~\cite{landau2018} (see also Ref.~\cite{beri2017} for a related discussion in the topological Kondo model). These different models: dynamical Coulomb blockade, Luttinger liquid, charge Kondo screening, have in fact been explored experimentally within a single setup~\cite{Iftikhar2018}. This underlines the experimental relevance of the fractional emitter we propose. Other sources of fractional emission have been predicted~\cite{berg2009} and measured~\cite{Kamata2014,inoue2014b} using coupled integer quantum Hall channels but with a non-rational charge.

The structure of the paper is as follows. Sec.\ref{sec-ballistic} introduces a simple model of ballistic edge channels in the quantum Hall effect covered by a gated island. The principles of fractionalization and anyonic statistics are discussed in this example geometry. They underpin the physics at play in the next Sections. In Sec.\ref{sec-emitter}, one of the channel is made quasi-ballistic by adding a weak backscattering point contact. It breaks the continuous flow of electrons and provides granularity to the charge transport. The current and noise show that a fractional charge is emitted, and Sec.\ref{sec-FCS} proves that it is poissonian by determining the full distribution of emission. The collider, or beam splitter mixing the outputs of two similar emitters, is investigated in Sec.\ref{sec-collider} by computing the cross-correlations. A discussion on the physical interpretation of the results follows. Sec.\ref{sec-summary} summarizes and concludes.

\section{Ballistic channels}\label{sec-ballistic}

\begin{figure}
\centering\includegraphics[width=\columnwidth]{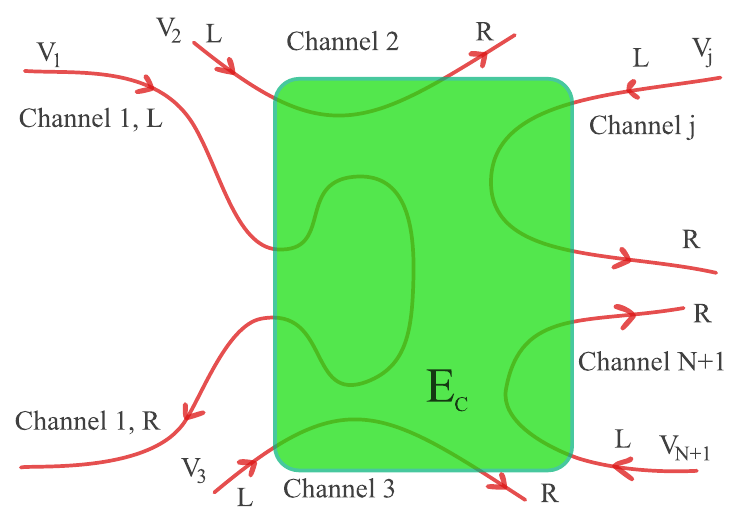}
\caption{Sketch of the apparatus fractionalizing electrons. $N+1$ propagating edge channels (in red) in the integer quantum Hall regime are partly covered by a gated island (in green) with charging energy $E_c$. The correspondence between incoming (L) and outgoing (R) chiral channels is inconsequential as the decoherence rate below the island is large. One electron arriving from channel $1$ (L) induces fractional charges $e/(N+1)$ in all $R$ channels. A voltage $V_j$ is applied upstream from channel $j$.\label{fig1} }
\end{figure}

\begin{figure}
\centering\includegraphics[width=\columnwidth]{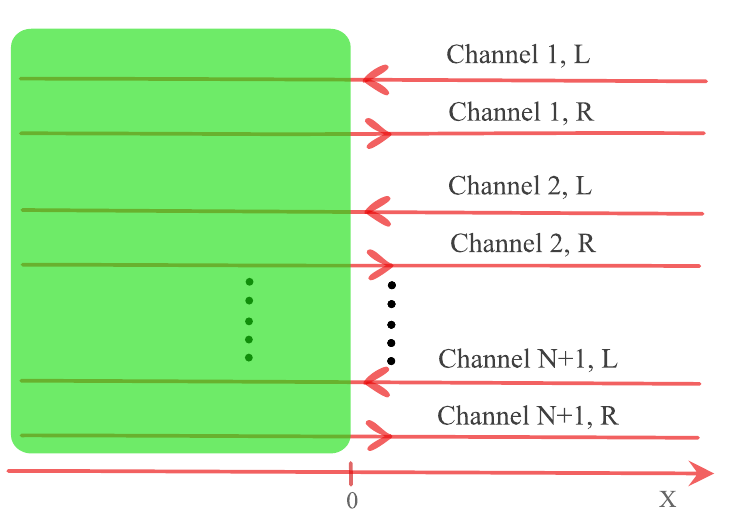}
\caption{Simplified description for the setup of Fig.\ref{fig1}. For each channel $j=1,\ldots,N+1$, the portion below the island in Fig.\ref{fig1} is replaced by two semi-infinite lines extending the incoming $L$ and outgoing $R$ parts. The axis $x$ labels the electron position. $x<0$ refers to the region inside the island and $x>0$ to outside. \label{fig2}}
\end{figure}

We lay out the principles for realizing an emitter of fractional charges in the integer quantum Hall effect. Under a strong magnetic field with filling $\nu=1$, electrons propagate along the edges of the two-dimensional sample. We first consider for simplicity a set of purely ballistic chiral channels, denoted by $j=1,\ldots,N+1$, with no tunneling between them. A metallic island is added which covers part of each channel as illustrated in Fig.\ref{fig1}. 
Using that electrons below the island region have a very short coherence time, we follow Refs.~\cite{Matveev1995,Furusaki1995b} and replace each edge path below the island by two decoupled semi-infinite chiral lines extending the incoming and outgoing chiral edges, as depicted in Fig.\ref{fig2}. Employing a bosonized description of the edge modes,  the Hamiltonian has the form $H_0 = \sum_j H_{0,j}$ with 
\begin{equation}
H_{0,j} = \dfrac{\hbar  v_F}{4\pi}\int_{-\infty}^{+\infty}\text{d}x\,[(\partial_x\phi_{L,j})^2+(\partial_x\phi_{R,j})^2]
\end{equation}
$L/R$ denotes the incoming/outgoing chiral fermions,  $x<0$ is the region below the island and $x>0$ the region outside the island.  The charging energy Hamiltonian $H_c = E_c \hat{N}^2$ involves the total charge below the island, given by
\begin{equation}\label{total-charge}
\hat{N} = \sum_{j=1}^{N+1} \int_{-\infty}^0 \left( \rho_{L,j} + \rho_{R,j} \right) = \frac{1}{2 \pi} \sum_{j=1}^{N+1} \left[ \phi _{R,j} (0) - \phi _{L,j} (0) )   \right]
\end{equation}
where the densities in the L and R modes are related to the bosonic fields through $\rho_{R/L,j} = \pm \frac{1}{2 \pi} \partial_x \phi_{R/L,j}$. $E_c  = e^2/2 C$ is the charging energy induced by the capacitance $C$ of the island to the ground. The Hamiltonian $H_0 + H_c$ is quadratic and can be solved. In order to write its explicit solution, it is convenient to rearrange the bosonic fields $\phi_{R/L,j}$, {\it i.e.} perform a rotation to define new chiral fields. We introduce the total charge field 
\begin{equation}
\tilde{\phi} _{R/L,1} = \frac{1}{\sqrt{N+1}} \sum_{j=1}^{N+1} \phi_{R/L,j},
\end{equation}
the other fields $\tilde{\phi} _{R/L,j}$ with $j\ge 2$ are defined to obtain a fully orthogonal set, details are deferred to Appendix~\ref{rotation}. The orthogonal transformation leaves the free part $H_0$ invariant when written in terms of the new bosonic fields. Since the total charge within the island is $\hat{N} = \sqrt{N+1} [ \tilde{\phi} _{R,1} (0) - \tilde{\phi} _{L,1} (0) )]/(2 \pi)$, all the new fields with $j \ne 1$  decouple from the island. This implies that an incoming electron $L$ has a decomposition over the new chiral modes ($j\ne 1$) that is completely absorbed by the island region, whereas electrons are independently emitted from the island to the outgoing channels $R$ ($j\ne 1$) . In other words, these chiral modes have a vanishing total charge and are therefore not sensitive to the island.

\begin{figure}
\centering\includegraphics[width=\columnwidth]{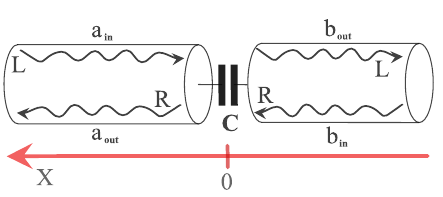}
\caption{Circuit description for the total charge bosonic mode: two transmission lines with impendance $R_q/2=h/2 e^2$ are in series with a capacitance $C/(N+1)$.  The other bosonic modes are fully absorbed by the island. The incoming and outgoing density waves are scattered by the capacitance with the unitary matrix~\eqref{sca1}. The left transmission line describes the incoming $L$ and outgoing $R$ bosonic channels outside the island, as displayed in Fig.\ref{fig1}. The right transmission line refers to the island. Note the opposite orientation of the $x$ axis in comparison with Fig.\ref{fig2}.\label{fig3}}
\end{figure}

The charging energy thus affects solely the total charge fields $\tilde{\phi} _{R/L,1}$. Interestingly, the corresponding Hamiltonian is exactly the same as the model of two semi-infinite transmission lines connected at the origin through a capacitance $C/(N+1)$, as depicted in Fig.\ref{fig3}. The transmission lines have each an impedance $R_q/2$ where $R_q = h/e^2$ is the unit of quantum resistance. One corresponds to the chiral edge states outside the island and is obtained by combining the corresponding $L$ and $R$ fields. The second transmission line describes the region inside the island, with both $L$ and $R$ components. Following standard techniques in the field of circuit quantum electrodynamics, one can develop an input/output formalism~\cite{safi1999,clerk2010}, by examining the Heisenberg evolution of the field operators, and describe the scattering of waves in the transmission lines by the central capacitance $C/(N+1)$. 
This is transparently discussed by expanding the fields over plane waves
\begin{equation}\label{expansion}
\tilde{\phi}_{L,1} (x,t) = \int_0^{+\infty} \frac{d \omega}{\sqrt{\omega}} \left( a_{in,\omega} e^{-i \omega (t+x/v_F)} + h.c. \right) 
\end{equation}
for $x>0$, where the Bose operators $a_{in,\omega}$ follow canonical quantization rules $[a_{in,\omega},a_{in,\omega'}^\dagger] = \delta(\omega-\omega')$. Similar expansions hold for $\tilde{\phi}_{L,1}$ ($x<0$), $\tilde{\phi}_{R,1}$ ($x<0$ and $x>0$) in terms of, respectively, $b_{out,\omega}$, $b_{in,\omega}$ and $a_{out,\omega}$. The subscript $in/out$ stands for the incoming/outgoing fields. The scattering by the capacitance at the origin is described by the unitary $S$ matrix
\begin{equation}\label{sca1}
\begin{pmatrix} 
a_{out,\omega} \\
b_{out,\omega}  
\end{pmatrix} = \begin{pmatrix}
r_0 & t_0 \\ t_0^* & r_0 
\end{pmatrix}   \begin{pmatrix} 
a_{in,\omega} \\
b_{in,\omega}  
\end{pmatrix} 
\end{equation}
The corresponding transmission and reflection coefficients are then functions of $\bar{\omega} =\pi\hbar \omega/(N+1) E_c$ where $\omega$ is the frequency of the wave,
\begin{equation}\label{sca2}
t _0(\bar{\omega}) = \frac{-i \bar{\omega}}{1- i \bar{\omega}},  \qquad r_0 (\bar{\omega}) = \frac{1}{1- i \bar{\omega}},
\end{equation}
as derived in Appendix~\ref{circuit}.
For a wave incoming in the $L$ channel, the transmission coefficient $t_0$ corresponds to electrons being transferred to the island, whereas the reflection coefficient $r_0$  describes backscattering at the entrance of the island. For energies or frequencies much smaller than the charging energy, the transmission coefficient vanishes and the wave is fully reflected $r_0=1$. In circuit words, the capacitance is replaced at small frequencies by an open link and the two transmission lines are disconnected. Eventually, restricting ourselves to low energies, which will be the case for the rest of this paper, we find the boundary condition
\begin{equation}\label{eq-bound}
\tilde{\phi}_{R,1} (x=0,t) = \tilde{\phi}_{L,1} (x=0,t)
\end{equation}
where the $R$ field is understood as an outgoing field and the $L$ field as an input field with respect to the scattering by the island entrance (capacitance) at $x=0$.

The relation~\eqref{eq-bound} has many interesting physical consequences. First, the complete reflection of the charge mode and the full transmission of the other chargeless modes explains the quantized reduction~\cite{Slobodeniuk2013} of the thermal conductance  observed in Ref.~\cite{Sivre2018}. The second aspect is the restoration of phase coherence~\cite{Clerk2001,lee2012} at low energy when a single chiral channel is present, or $N=0$. In bosonization, the boson fields describe the phase of the fermions. Thus the boundary condition Eq.~\eqref{eq-bound} means that electrons escaping the island are phase-coherent~\cite{aleiner1998,Filippone2020} with the electrons impinging the island for $N=0$. In the case of a single edge mode, the total charge mode coincides with the only mode available. Physically, an electron entering the island violates energy conservation because it changes the charging energy. It must therefore be compensated immediately, at least on time scales larger than $\hbar/E_c$, by an electron leaving the island. The restored phase coherence is a result of a Fermi liquid mechanism~\cite{filippone2011,filippone2012} as the phase space available for particle-hole excitation vanishes at low energy. The resulting phase coherence in the electron transfer across the island, also coined as electron teleportation~\cite{Fu2010,lee2012}, has been successfully measured~\cite{Duprez2019b,Idrisov2018} by embedding the island and edge channel within a Mach-Zender interferometer. We emphasize that the phase coherence is also responsible for a quantized resistance in the quantum RC circuit~\cite{mora2010}. Phase coherence and Fermi liquid behaviour are lost as soon as we have more than one channel $N>0$. 

In this paper, we will focus on another important consequence of Eq.~\eqref{eq-bound}, namely charge fractionalization~\cite{lee2020,andreev2001,Idrisov2019}. Physically, it arises naturally from the above considerations. One electron coming from, say channel $1$, has a linear decomposition over the new chiral modes. As only the total charge mode is not absorbed by the island, the electron entering the island triggers the emission of fractional charges $e/(N+1)$ in all available output channels. This can be seen by applying a voltage $V$ to channel $1$ and computing the currents. In the $L$ region before the island, the Bose field is given by $\phi_{L,1} - e V t/\hbar$, where $\phi_{L,1}$ is given by Eq.~\eqref{expansion} and the field $a_{in,\omega}$ is characterized by a thermal Bose occupation and $\langle a_{in,\omega} \rangle =0$. The other $L$ fields ($j\ne1$) have no applied potential, hence $\langle \phi_{L,j} \rangle=0$. The current operator {\it towards} the island is given by $\hat{I}_{L,1} = -(e/2 \pi) \partial_t \phi_{L,1}$, where $e$ is the electric charge, and its expectation reads
\begin{equation}
\langle \hat{I}_{L,1} \rangle = \frac{e^2}{h} \, V.
\end{equation}
The output field $\phi_{R,1}$ in channel $1$ can be written in terms of the new chiral fields
\begin{equation}\label{newchiral}
\phi_{R,1} = \frac{1}{\sqrt{N+1}} \, \tilde{\phi}_{R,1} + \sqrt{\frac{N}{N+1}} \, \tilde{\phi}_{R,2}. 
\end{equation}
The new field with index $2$ is chargeless and thus independent from the incoming electrons. It feels no voltage bias and has simply $\langle\tilde{\phi}_{R,2} \rangle=0$. In contrast, the total charge field with index $1$, thanks to Eq.~\eqref{eq-bound}, sees a non-zero potential bias, or more explicitely, Eq.~\eqref{eq-bound} becomes
\begin{equation}
\tilde{\phi}_{R,1} = \frac{1}{\sqrt{N+1}} \left( \phi_{L,1} - e V t/\hbar + \sum_{j=2}^{N+1} \phi_{L,j} \right).
\end{equation}
Computing the output ($R$) current in channel $1$, we obtain
\begin{equation}\label{output1}
\langle \hat{I}_{R,1} \rangle = - \frac{e}{2 \pi} \langle \partial_t \phi_{R,1} \rangle  = \frac{e^2}{h} \, \frac{V}{N+1}
\end{equation}
showing the charge fractionalization. Computing the current in all other output channels reproduces Eq.~\eqref{output1}. Despite splitting the current, this is not completely what we want for an emitter because the flow of electrons is continuous and no granularity is visible. Computing the noise in the output channel, one easily obtains that it vanishes at zero temperature: there is no shot noise. This will be remedied in the next Section by adding a weak quantum point contact to channel $1$. In the general case where the potential bias applied to channel $j$ is $V_j$, the output current has the form
\begin{equation}
\langle \hat{I}_{R,1} \rangle  = \frac{e^2}{h} \, \frac{\sum_j \, V_j}{N+1},
\end{equation}
and it vanishes when the sum over all biases is zero. This is because, in that case, the potentials excite only chiral chargeless modes which are absorbed by the island and no electron leave the island.

In the fractional quantum Hall effect, the presence of fractional charges is related to the anyonic phase shift from Laughlin's gauge argument. Here, we will show that the simple input/output approach supporting charge fractionalization also contains the seeds for anyonic statistics. In bosonization, the fermionic operators have a non-linear relation to the bosonic fields
\begin{equation}\label{bosonization}
\psi_{R/L,j} (x) =  \sqrt{\dfrac{D}{2\pi \hbar v_F}} e^{i\phi_{R/L,j}(x)}
\end{equation}
$D$ is a high-energy cutoff (bandwidth) for the chiral edge electrons necessary to regularize the theory. It will disappear in the evaluation of physical quantities. We consider the tunneling operator
\begin{equation}\label{tunneling1}
{\cal T} (t) = \psi^\dagger_{R,1} (0,t) \psi_{L,1} (0,t) 
\end{equation}
in the Heisenberg representation, taking channel $1$ without loss of generality. Using the bosonization formula Eq.~\eqref{bosonization}, and the new chiral fields with Eq.\eqref{newchiral}, we obtain
\begin{equation}\label{tunneling2}
{\cal T} (t) =  \frac{D}{2\pi \hbar v_F} e^{i \left[ \frac{1}{\sqrt{N+1}} \, (\tilde{\phi}_{L,1} - \tilde{\phi}_{R,1} )  + \sqrt{\frac{N}{N+1}} \, ( \tilde{\phi}_{L,2} - \tilde{\phi}_{R,2} ) \right]}
\end{equation}
where we use the notation $\tilde{\phi}_{R/L,j}= \tilde{\phi}_{R/L,j}(0,t)$. The boundary condition~\eqref{eq-bound} imply that the first term in the exponential should vanish. In practice, to properly account for the renormalization of high-energy ($\sim E_c$) modes, we integrate in Appendix~\ref{appen-bosonization} Eq.~\eqref{tunneling2} over the massive fields $\tilde{\phi}_{L/R,1}$ using the scattering matrix from Eqs.\eqref{sca1} and~\eqref{sca2}. Exploiting the Baker-Campbell-Hausdorff formula, we find the low-energy tunneling operator
\begin{equation}\label{tunneling3}
{\cal T} (t) = {\cal N} e^{i\sqrt{\frac{N}{N+1}} \, ( \tilde{\phi}_{L,2} - \tilde{\phi}_{R,2} )}
\end{equation}
with the prefactor $\cal {N} =$ $\frac{D}{2\pi \hbar v_F}\left(\frac{e^\gamma (N+1)E_c}{\pi D}\right)^{1/(N+1)}$ where $\gamma=0.5772$ is  Euler's constant. After this integration, the local tunneling operator acquires the anomalous dimension $\nu = N/(N+1)$, $\langle {\cal T}^\dagger (t) {\cal T} (0) \rangle \sim 1/t^{2 \nu}$. It also obeys fractional statistics with itself, $  {\cal T} (t) {\cal T} (0) =  {\cal T} (0) {\cal T} (t) e^{2i\pi\nu\,\text{sgn}(t)}$, and with the fermionic (outgoing) field,   
\begin{equation}\label{mutualstatistics}
{\cal T} (t) \psi_{R,1} (0) =  \psi_{R,1} (0) {\cal T} (t) e^{- i \pi \nu\,\text{sgn}(t)}.
\end{equation}

\section{Fractional emitter}
\label{sec-emitter}

\begin{figure}
\centering\includegraphics[width=\columnwidth]{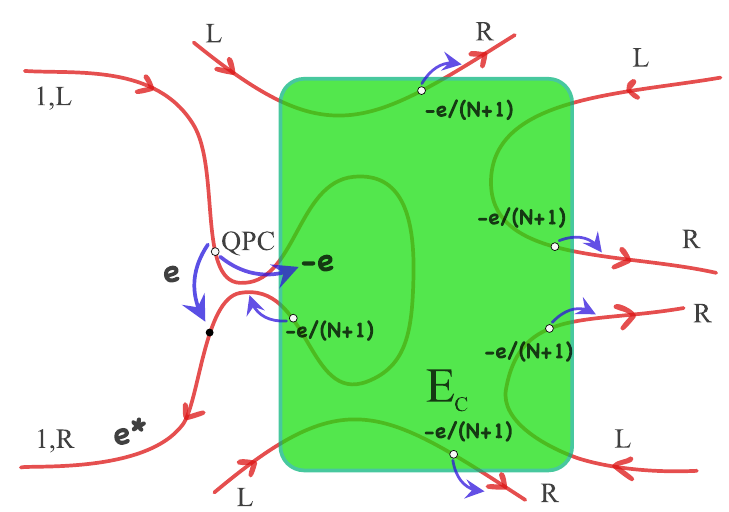}
\caption{Principle of the fractional emitter. We add a QPC to Fig.\ref{fig1}. The QPC directly couples, with a small amplitude $r$, the two parts ($L$ and $R$) of channel $1$. When a single electron (black dot) tunnels via the QPC from $L$ to $R$, it leaves a hole (white dot) of charge $-e$ behind. The hole enters the island and fractionalizes into a multiplet of charges $-e/(N+1)$ hopping out of the dot. The net resulting charge $e^* = \nu e$, see Eq.~\eqref{charge}, in channel $1$ and $-\frac{e}{N+1}$ in the other channels. \label{fig4}}
\end{figure}

The flow of electrons is continuous for ballistic channels and, in the absence of electron backscattering, the granularity of the electric charge is not apparent. It can be seen as a consequence of the quadratic Hamiltonian with bosonic variables, although the corresponding fermionic Hamiltonian is interacting, whereas the current operator is linear in the Bose fields. In this configuration, there can be no shot noise revealing elementary charges. We thus realize our emitter by adding a weakly coupled quantum point contact in channel $1$, as depicted in Fig.\ref{fig4}.  The resulting fractional charge emission is described below.

\subsection{Setup and fractional charges}

The quantum point contact is located close to the island, at $x=0$ in channel 1.
The corresponding Hamiltonian is
\begin{equation}
H_{BS} = \hbar v_F r \left( {\cal T} + {\cal T}^\dagger \right),
\end{equation}
with the small backscattering coefficient $r \ll 1$. The tunneling operator has been introduced in Eq.~\eqref{tunneling1} and the total Hamiltonian for the emitter is $H_0 + H_c + H_{BS}$. In order to avoid a continuous flow of electrons and to focus on the fractional emitted charges, we take the symmetric configuration where $\sum_j V_j = 0$ so that the island emits zero charge in the absence of $H_{BS}$ and for vanishing temperature.

We are interested in computing the properties of the output current $\hat{I}_{R,1}$, defined in Eq.~\eqref{output1}, which characterizes the electrons leaving the island region, either because they were expelled from the island or because they are scattered electrons coming from the chiral branch $L$ of the same channel $1$. To proceed with its calculation, we use an interaction representation with respect to the backscattering term $H_{BS}$. We thus time-evolve the bosonic fields according to $H_0+H_c$ as discussed in the previous Section with the input/output approach, and dress the current with the also time-evolved backscattering term. The starting point is the formula
\begin{equation}\label{interaction}
\hat{I}_{R,1} (t) = U_I^\dagger (t) \hat{I}^{(I)}_{R,1} (t) U_I(t) 
\end{equation}
with the evolution operator $U_I (t) = e^{-\frac{i}{\hbar} \int_{\infty}^t d t' H_{BS} (t')}$ and the current operator in the interaction representation is
\begin{equation}\label{current1}
\hat{I}^{(I)}_{R,1} (t) = - \frac{e}{2 \pi}\sqrt{\frac{N}{N+1}} \partial_t \tilde{\phi}_{R,2} (t). 
\end{equation}
We have used  Eq.~\eqref{newchiral} and dropped the $\tilde{\phi}_{R,1}$ contribution since this mode is massive at energies well below the charging energy $E_c$. Accordingly, we take the low-energy expression~\eqref{tunneling3}, or
\begin{equation}\label{tunneling4}
{\cal T} (t) = {\cal N} e^{- i h(t) - i e V t/\hbar} 
\end{equation}
with the effective voltage $V = \frac{2 N V_1}{N+1}$ and the notation $h =\sqrt{\frac{N}{N+1}} \, ( \tilde{\phi}_{R,2} - \tilde{\phi}_{L,2} )$. We stress again that the sum over all bias voltages is chosen to vanish. 

We treat $H_{BS}$ perturbatively and expand Eq.~\eqref{interaction} to second order in $r$, $\hat{I}_{R,1} = \hat{I}^0 + \hat{I}^1+\hat{I}^2$, with $\hat{I}^0 = \hat{I}^{(I)}_{R,1}$ and
\begin{subequations}\label{currents}
\begin{align}
\hat{I}^1 &= -i r v_F \, e^* [{\cal T}(t)-{\cal T}^\dagger(t)]   \\
\hat{I}^2 &= e^* (r v_F)^2\int_{-\infty}^{+\infty}\text{d}t'\,[{\cal T}^\dagger(t'),{\cal T}(0)]
\end{align}
\end{subequations}
where, anticipating the final outcome, we introduced the fractional charge
\begin{equation}\label{charge}
e^* = \nu \, e \qquad \nu = \frac{N}{N+1}.
\end{equation}
We are now in a position to evaluate the current $I$ and the (zero-frequency) current noise $S$ by taking the expectation of the operator expression from Eq.~\eqref{currents}. We obtain
\begin{equation}\label{current}
I = e^* r^2 \frac{e V}{\hbar}  \left( \frac{\tilde{E}_c}{\nu e V} \right)^{2 (1-\nu)} \frac{ \Gamma(1-2 \nu) \sin 2 \pi \nu}{2 \pi^2}
\end{equation}
for the current and 
\begin{equation}\label{noise-result}
S = e^* I
\end{equation}
for the noise. We define the renormalized charging energy $\tilde{E}_c = N E_c e^\gamma/\pi$~\footnote{$\tilde{E}_c$ can also be described as arising from the capacitance $C$ with the resistance $R_q/N$ built from all parallel ballistic channels of resistance $R_q = h/e^2$.}. 
The result~\eqref{current} is limited to weak backscattering $r^2 (E_c/e V)^{2-2 \nu} \ll 1$~\footnote{The tunneling operator~\eqref{tunneling4} is a relevant operator in the renormalization group sense. The QPC fully reflects incoming electrons at very low energy and an intermediate energy regime is necessary to validate a perturbative calculation.}. In computing the current, only $\hat{I}^2$ has a non-zero quantum average whereas only the product $\langle \hat{I}^1 \hat{I}^1 \rangle$ contributes to the noise at zero frequency. When the number of channels is infinite $N \to + \infty$, we recover a non-interacting backscattering, $I = r^2 e^2 V/h$ and $\nu =1$, indicating a dilution of the charging energy into the many available channels.

The result for the Fano factor $F=S/I=e^*$ suggests the emission of fractional charges by the island and the quantum point contact into the edge channel $R$. A simple physical picture can be given. For vanishing $r$, there is no current or noise in the lead $R$ so that only backscattering processes controlled by $r$ triggers the emission of charges. Since $r$ is small, one expects the train of emitted charges to be dilute and the corresponding statistics to be poissonian. We note that the effective charge $e^*$ given in Eq.~\eqref{charge} is slightly different from the fractionalization mentioned in the previous Section. There are indeed two processes for the charge transfer: (i) a single electron is scattered from the $L$ to the $R$ branch in channel $1$, (ii) the process (i) leaves a hole in the $L$ branch which then enters the island with a charge $-e$ and must be automatically compensated by holes with fractional charges $-e/(N+1)$  expelled in all channels. The two processes (i) and (ii) add coherently and trigger the emission of the fractional charge $e^* = e - e/(N+1)$. This coherent sum is in contrast with the Fano factors measured in Kondo screened islands where poissonian processes add incoherently~\cite{sela2006,mora2009,Ferrier2016,ferrier2017}.

The physical picture drawn here implies that the processes (i) and (ii) also trigger the poissonian emission of fractional charges $-e/(N+1)$ into all the channels with $j \ge 2$. This is verified in Appendix~\ref{alternative} by computing the corresponding current and noise.

\subsection{Poissonian emission of fractional charges} \label{sec-FCS}

Although the physical picture for the charge emission is more than reasonable, we confront it by computing the full counting statistics (FCS)~\cite{levitov1996} of our fractional emitter. The FCS is determined by the characteristic function~\cite{esposito2009}
\begin{equation}\label{charac}
\ln\chi (\lambda) =  \langle
e^{i\lambda \hat{N}_{R,1}(t)}e^{-i\lambda \hat{N}_{R,1}(0)}\rangle
\end{equation}
where $\hat{N}_{R,1} (t) = \int^t d t' \hat{I}_{R,1} (t')/e$ is the operator giving the emitted charge taken at position $x=L$, sufficiently far from the island. Using the current expression, it also relates directly to the outgoing bosonic field, $\hat{N}_{R,1} (t) =  -\phi_{R,1} (t)/2 \pi$.
From $\chi(\lambda)$ we can retrieve  all cumulants of the emitted charge. After standard manipulations, the generating function can be written as
\begin{equation}\label{fcs}
\ln\chi (\lambda) =  \langle T_K e^{- \frac{i}{\hbar} \int_K d t H_{ \pm\lambda} (t)} \rangle
\end{equation}
where the time integration follows the Keldysh contour with a forward $+$ and a backward branch $-$.  The Hamiltonian is dressed by a counting field $\lambda$ with $H_\lambda = H  - \hbar\lambda \hat{I}_{R,1}/(2e)$. $\lambda$ is non-zero only in the time interval $[0,T]$ and has opposite signs on the two Keldysh branches. Similarly to Sec.\ref{sec-emitter}, we work in the interaction representation and include the counting field $\lambda$ into the Heisenberg evolution for the bosonic fields $\phi_{R/L,j}$. As detailed in Appendix~\ref{counting-field}, we simply obtain that the rotated chiral fields $\hat{\phi}_{R/L,1}$ are both shifted by -$\eta \frac{\lambda}{2 \sqrt{N+1}}$ where $\eta = \pm$ is the Keldysh contour index, whereas $\hat{\phi}_{R,2}$ is shifted by -$\eta \frac{\lambda}{2} \sqrt{\frac{N}{N+1}}$ and  $\hat{\phi}_{L,2}$ is left invariant. The expression obtained for the generating function is
\begin{equation}\label{fcs1}
\ln\chi (\lambda) =\langle T_K e^{- i v_F r \int_K d t \left [ {\cal T} (t)  e^{i \eta \lambda \nu/2} + {\cal T}^\dagger (t)  e^{-i \eta \lambda \nu/2} \right]}
 \rangle
\end{equation}
where the tunneling term ${\cal T} (t)$, see Eq.~\eqref{tunneling4}, is dressed by the counting field. Expanding to second order in $r$ results in a double time integral. Switching to the time difference and average, the time difference variable can be integrated by assuming a sufficiently large time interval $T$, with the result
\begin{equation}\label{charac2}
\ln\chi (\lambda) \simeq -\dfrac{I}{e^*}|t|\,
\left(1-e^{i\lambda\nu}\right).
\end{equation}
Remarkably, this result is exactly the generating function describing a poissonian emission by the fractional emitter with the effective charge $e^* = \nu e$, in agreement with the physical argument previously presented.

\section{Collider of fractional beams}\label{sec-collider}

\begin{figure}
\centering\includegraphics[width=\columnwidth]{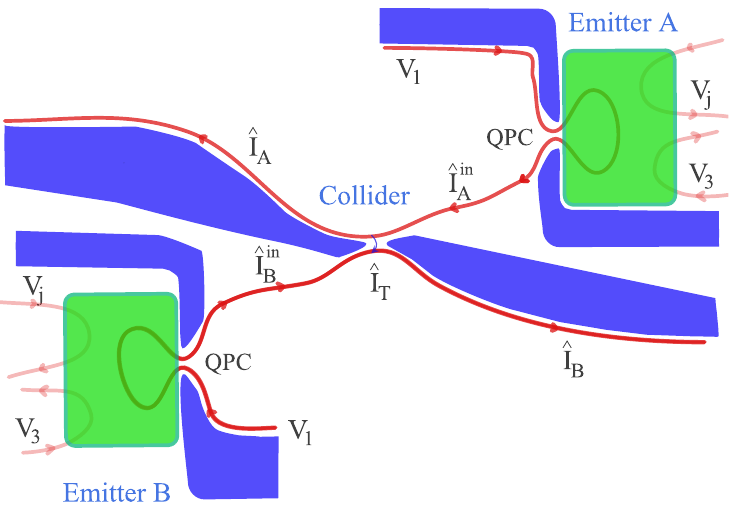}
\caption{Sketch of the full collider device probing the noise properties of fractional beams. The two emitters, named $A$ and $B$, are realized according to Fig.~\ref{fig4} \cm{with $N$ ballistic channels each}. The collider acts as a beam splitter for the output channels $R$ of both emitters. We denote $L$ the distance between each emitter and the collider. \cm{The blue regions represent the gates defining the three quantum point contacts.} \label{fig5}}
\end{figure}

So far we have shown how to realize an emitter of dilute fractional charges by using an island and quasi-ballistic edge channels in the integer quantum Hall regime. The emitter also entails exchange relations with a statistical anyonic phase. We now consider two such fractional emitters, which we will denote with the letters $A$ and $B$, and study the regime of $e^2 V / h \gg I_{A/B}$ in which the fractional emitters send dilute beams to a collider setup. Here $I_{A/B} = \langle \hat{I}^{\rm in}_{A/B}\rangle$ is the current~\eqref{current} generated by the emitter $A/B$.
The collider, a quantum point contact positioned far enough from the islands, acts as a beam splitter which partitions the incoming fractional charges. The corresponding device is depicted in Fig.~\ref{fig5}.

We denote $t_P$ and $r_P$ the transmission and reflection amplitude coefficients for the collider. We use the notation $h_{A/B}$ for the field entering the tunneling in Eq.~\eqref{tunneling4} and $\psi_{A,B}$ ($\phi_{A,B}$) for the chiral fields $\psi_{R,1}$ ($\phi_{R,1}$) at the output of each emitter.

\subsection{Collision of free electrons}

Before delving into the fractional case, we first review the collision with free fermions. The cross-correlation at the output of the beam splitter follows from Landauer-B\"uttiker formalism\cite{Blanter2000},
\begin{equation}\label{landauer}
\begin{split}
 \langle \delta \hat{I}_A \delta \hat{I}_B \rangle  = T_P (1-T_P) \frac{e^2}{h} \int d  \varepsilon 
\Big[ f_A (1-f_A) \\
 + f_B (1-f_B) - f_A (1-f_B) - f_B (1-f_A) \Big] 
\end{split}
\end{equation}
 or $\langle \delta \hat{I}_A \delta \hat{I}_B \rangle  = - T_P (1-T_P) \frac{e^2}{h} \int d  \varepsilon (f_A-f_B)^2$, where $\delta \hat{I}_A = \hat{I}_A - \langle \hat{I}_A  \rangle$ and $T_P = |t_P|^2$. $f_{A/B} (\varepsilon)$ is the energy distribution of the beam incoming from emitter $A/B$. The derivation of Eq.~\eqref{landauer} relies crucially on a gaussian density operator for the use of Wick's theorem. It is the case for instance in the out-of-equilibrium situation where the incoming beam is obtained by coherently mixing, with an upstream quantum point contact, two fermionic populations with different chemical potentials. 

For identical distributions $f_{A/B}$, the cross-correlation~\eqref{landauer} exactly vanishes. This cancellation can also be understood by invoking current conservation~\cite{Blanter2000}. The noise before the collider is equal to the noise after, including the cross-correlation,
\begin{equation}
\langle \delta \hat{I}^{\rm in}_A \delta \hat{I}^{\rm in}_A \rangle 
+ \langle \delta \hat{I}^{\rm in}_B \delta \hat{I}^{\rm in}_B \rangle = 
\langle \delta \hat{I}_A \delta \hat{I}_A \rangle 
+ \langle \delta \hat{I}_B \delta \hat{I}_B \rangle + 2
\langle \delta \hat{I}_A \delta \hat{I}_B \rangle.
\end{equation}
Since the fermionic fields after the collider are linear combination of the incoming fields, {\it i.e.} $\psi_A = t_P \psi^{\rm in}_A + r_P \psi^{\rm in}_B$, they inherit the same energy distribution, hence  $\langle \delta \hat{I}^{\rm in}_{A/B} \delta \hat{I}^{\rm in}_{A/B} \rangle = \langle \delta \hat{I}_{A/B} \delta \hat{I}_{A/B} \rangle$, such that the cross-correlation has to be zero. The lesson here is that negative cross-correlations for identical colliding beams must be a signature of interaction: electrons are dressed with clouds of electron-hole pairs as it is the case for a fractional quasiparticle. After the collision, the sum of noises in the outgoing channel is increased by $-\langle \delta \hat{I}_A \delta \hat{I}_B \rangle$.

An alternative picture can be proposed for the cross-correlation noise~\cite{rosenow2016}. In Eq.~\eqref{landauer}, the first two terms are proportional to the noise present in each dilute beam. As part of the incoming beam is reflected by the collider, cross-correlation measures some of its shot noise. The third and fourth terms are partition noise at the quantum point contact as electrons leave holes behind and give negative cross-correlations. The Fermi statistics plays a role in partition noise when electrons arrive simultaneously: they cannot exit the collider on the same edge  because of the Pauli principle. This has the effect of reducing the magnitude of partition noise. As a consequence, violation of Pauli by anyonic statistics increases partition noise and tilts the balance towards an overall negative cross-correlation in a way that depends on the statistical phase.

This is why the collider setup has been expected to be useful for detecting anyonic statistics.
As shown below, however, this effect is masked by another striking partition mechanism that involves double exchange (braiding) between a fractional charge and an electron, hence, twice the mutual fractional statistical phase in Eq.~\eqref{mutualstatistics}.
This mechanism has no counterpart for free electrons.

\subsection{Tunneling current and noise at collider}\label{sec-collider2}

We now turn to our fractional emitters and examine the mixing of their output beams as represented in Fig.~\ref{fig5}. 
In order to evaluate the resulting noise properties, we treat the electron tunneling at the collider (the central QPC) perturbatively with
\begin{equation}
H_T = \hbar v_F  t_P  \psi_A^\dagger\psi_B + h.c.
\end{equation}
and the corresponding current, from $A$ to $B$, is described by $\hat{I}_T = i e v_F t_P ( \psi_A^\dagger\psi_B-h.c.)$. 

The tunneling current $I_T = \langle \hat{I}_T \rangle$
\begin{eqnarray} \label{ITcurrent} 
I_T   = e v_F^2 T_P
\int_{-\infty}^{+\infty}\text{d}t\langle [ \psi_A^\dagger(t)\psi_B(t),\psi^\dagger_B(0)\psi_A(0) ] \rangle  \quad \quad
\end{eqnarray}
involves the time correlation functions for the fields $\psi_{A/B}$ and  the transmission probability $T_P = |t_P|^2$ of the QPC.
Taking advantage of the linear relations between the fermionic operators before and after the QPC collider~\cite{Blanter2000}, the average tunneling current  between the $A$ and $B$ channels is readily obtained
\begin{equation}\label{tunneling-current}
I_T =  T_P ( I_A - I_B)    
\end{equation}
where $I_{A/B} = \langle \hat{I}^{\rm in}_{A/B}\rangle$ is the current~\eqref{current} generated by the emitter $A/B$. 
As expected by symmetry, it vanishes when the emitters are identical. 

The noise of the tunneling current is written as
\begin{equation} \label{noise}
\frac{\langle \delta \hat{I}_T \delta \hat{I}_T\rangle}{(ev_F)^2 T_P}  =
\int_{-\infty}^{+\infty}\text{d}t\langle \big\{\psi_A^\dagger(t)\psi_B(t),\psi^\dagger_B(0)\psi_A(0) \} \rangle. 
\end{equation}
The noise is determined by the long time behavior for which a non-perturbative evaluation with respect to $r$ or $I_{A/B}$ is necessary.
The time correlation functions may be written in the Keldysh framework as
\begin{equation}\label{twopoint}
\langle \psi_A(t) \psi^\dagger_A(0)\rangle= \langle T_K \psi_A(t^-)\psi^\dagger_A(0^+)
e^{-\frac{i}{\hbar} \int_K d t' H_{BS} (t') } \rangle 
\end{equation}
where the expectation value  and the time evolution of $H_{BS} (t)$ are governed by the Hamiltonian $H_0+H_c$. The time $t$ (resp. $0$) is taken on the backward (resp. forward) Keldysh path. The field $\psi_A$ is taken at the position $x=L$ where $L$ is large enough, while the tunneling $H_{BS} (t)$ in the emitter $A$ is at $x=0$. The integral in Eq.~\eqref{twopoint} is dominated by times, irrespective of the Keldysh branch, between $-L/v_F$ and $-L/v_F +t$, which are both smaller than $0$ and $t$ for sufficiently large $L$. $L/v_F$ is the time needed for an electron to propagate between the island and the collider. 

Expanding Eq.~\eqref{twopoint} to second order in $r$, one finds the perturbed correlation function ($t>0$)
\begin{equation}
\langle \psi_A(t)\psi^\dagger_A(0)\rangle= \frac{1}{2 \pi v_F}  \frac{1}{a+i  t}  \left[ 1 - \left(\frac{v_F r {\cal N}}{a^{-\nu}} \right)^2 {\cal K} (t)\right], 
\end{equation}
with the short-time cutoff $a=\hbar/D$ and
\begin{equation}\label{functionK}
\begin{split}
 {\cal K} (t) = \sum_{\eta_{1/2}} &\eta_1\eta_2
\int_0^{t} d t_1\, \int_0^{t} d t_2\, \frac{e^{i e V (t_1-t_2)/\hbar}}{[a+i (t_1-t_2)\,\chi_{1,2}]^{2 \nu}} \\[2mm]
& \left(\frac{a- i t_2 \,\eta_2}{a-i t_1 \,\eta_1} \right)^{\nu}
\left(\frac{a+ i  (t-t_1)\,\eta_1}{a+i (t-t_2)\,\eta_2} \right)^{\nu},
\end{split}
  \end{equation}
where the time integrals have been shifted to absorb the transit time $L/v_F$. Here $\chi_{1,2} = \eta_2$ ($-\eta_1$) for $t_1 > t_2$ ($t_2 > t_1$). The integral is dominated by $t_1 \simeq t_2$. Switching to the time difference and time average variables, similarly to the computation done for the FCS, one performs the integration over the difference and find~\cite{Note4} (for $t>0$)
\begin{equation}\label{twopoint2}
\langle \psi_A(t)\psi^\dagger_A(0)\rangle \simeq \frac{1}{2 \pi v_F}  \frac{1}{a+i t}  \left[ 1 -  \frac{I_A}{e^*} \left( 1 - e^{- 2 i \pi \nu}  \right)  \,  t \right],
\end{equation}
where $I_A$ is the current~\eqref{current} generated by the emitter $A$.  For $t<0$, we can use the  relation $\langle \psi_A(t)\psi^\dagger_A(0)\rangle=\langle \psi_A(-t)\psi^\dagger_A(0)\rangle^*$. 

The finding Eq.~\eqref{twopoint2} shows that the perturbative correction dominates the long-time asymptotic and a non-perturbative calculation (in $r$) is necessary. It can be done in the present case order by order~\cite{Note4}  and the resummation leads to 
\begin{equation}\label{twopoint3}
\langle \psi_A(t)\psi^\dagger_A(0)\rangle \simeq \frac{1}{2 \pi v_F}  \frac{ e^{-  \frac{I_A}{e^*} \left( 1 - e^{- 2 i \pi \nu}  \right)  \,  t}} {a+i t}. 
\end{equation}
Equations~\eqref{twopoint} and  \eqref{twopoint3} were derived in the long-time limit, largely exceeding the time-spread of an electron wave packet $t \gg h/e V$~\footnote{In realistic situations, the temporal width of a fractional charge $\nu e =  e N/(N+1)$ is determined by $h/(eV)$ and twice the temporal distance between 
 its two constituents, an electron $e$ and a fractional hole $- e / (N+1)$; in this work, this temporal distance is set to zero for simplicity, as in Eq.~\eqref{tunneling1}. The temporal width of fractional charges $- e / (N+1)$ in the other output channels $j \ge 2$ is determined solely by $h/(eV)$.}. Therefore, the short-time cutoff $a$ should be replaced by $h/e V$ in Eq.~\eqref{twopoint3}.
The exponential decay occurs over the time scale $ e/I_A \gg h/ e V$. This last separation of time scale is equivalent to having  $r^2 (E_c/e V)^{2-2 \nu} \ll 1$ controlling the perturbative calculation of last Section, see Eq.~\eqref{current}. Exchanging $\psi$ by $\psi^\dagger$ in Eq.~\eqref{twopoint3}, one obtains the same expression with $\nu \to - \nu$.

The exponential term in Eq.~\eqref{twopoint3} comes from the double exchange between fractional charges and an electron~\footnote{See Supplemental Material where the analytical computations in the Keldysh framework are detailed. The field correlator, including its non-perturbative resummation, is derived.}, see Eq.~\eqref{mutualstatistics}, hence, from their fractional mutual statistics.
It is the average of the double exchange phase $e^{-i 2 \pi m \nu}$ over the Poissonian distribution $p(m)$  of $m$ fractional charges in the time interval $t$,
\begin{equation} \label{poissonian_braiding}
 e^{-  \frac{I_A}{e^*} \left( 1 - e^{- 2 i \pi \nu}  \right)  \,  t} = \langle e^{-i 2 \pi  m \nu} \rangle = \sum_m p(m) e^{-i 2 \pi  m \nu},
\end{equation}
where $p(m) = (\bar{m}^m / m!)e^{- \bar{m}}$ and $\bar{m} = I_A t / e^*$ is the average number of fractional charges in the interval.
Therefore, this expression coincides with the FCS characteristic function for $\lambda = 2 \pi$ given Eq.~\eqref{charac2}. This identification follows readily~\cite{rosenow2016} from the bosonization formula~\eqref{bosonization} for $\psi_A$, the generating function Eq.~\eqref{charac} and the charge $\hat{N}_A = -\phi_A/2 \pi$. 

The combination $(1- e^{-2i \pi \nu})$ appears to each order in perturbation theory for the correlator $\langle \psi_A (t) \psi^\dagger_A(0) \rangle$. Its physical meaning will be clarified later in Sec.~\ref{section_onesideinjection}. It vanishes in the free electron case of $\nu = 1$ where the double exchange is trivial. In this case, including the subleading corrections to Eq.~\eqref{twopoint3}, given in Eq.~\eqref{longtime} in Appendix \ref{Appendix_detailed_calculation}, is necessary to retrieve the free fermion form.

The field correlation function~\eqref{twopoint3} is finally used to
compute the partition noise in Eq.~\eqref{noise}. After performing the time integral, we arrive at the dominant contribution to the tunneling noise,
\begin{equation}\label{noise2}
\frac{\langle \delta \hat{I}_T \delta \hat{I}_T \rangle}{e^2 T_P} \simeq \frac{2 I_+ \sin^2 \pi \nu}{\pi^2 e^*} \ln\left(\frac{e^* e V/h }{2|\sin \pi \nu| \tilde{I} (\nu) }\right),
\end{equation}
where $\tilde{I} (\nu) = \sqrt{I_+^2 \sin^2 \pi \nu + I_-^2 \cos^2 \pi \nu}$ and $I_\pm = I_A \pm I_B$.
The dependence on $\nu$ originates from the double exchange between fractional charges and an electron.

\begin{figure*}[t!]
	\centering
	\includegraphics[width = 0.9\textwidth]{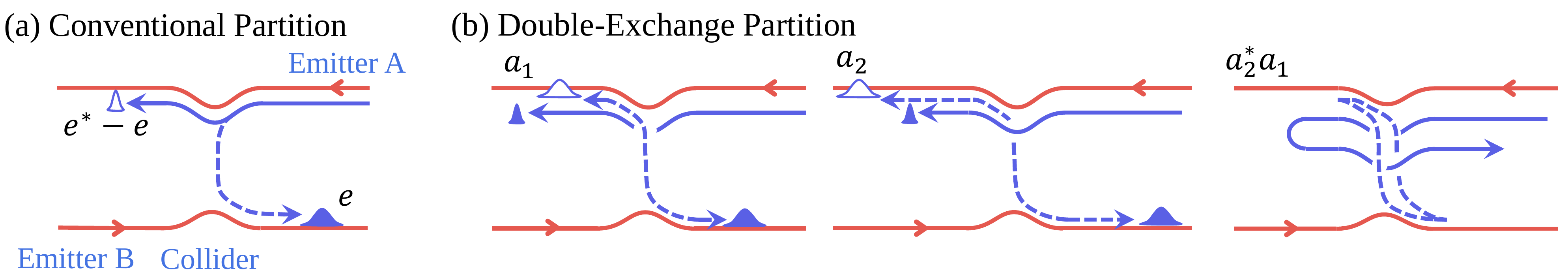}
	\caption{Partition at $I_A \ne 0$, $I_B = 0$, and $r \ll 1$.
(a) Conventional partition. When a fractional charge $e^*$ arrives at the collider QPC, an electron tunneling (dashed) happens from the channel A to B. Then a charge $e^*-e$ (thin empty packet) moves along the channel A, and an electron (thick filled) moves along B. Another partition happens with electron tunneling from B to A (not shown).
(b) Double-exchange partition. In a subprocess $a_1$ [resp. $a_2$], a pair of an electron (thick filled) and a hole (thick empty) is excited at the collider QPC after [resp. before] a fractional charge $e^*$ (thin filled) passes the QPC. Crossing of two trajectories  is time ordered such that the later trajectory is drawn on top of the earlier. The interference $a_2^* a_1$ between the two subprocesses involves a double exchange (braiding) between $e^*$ and $e$. Another partition happens with a pair excitation of an electron on the channel A and a hole on B at the QPC (not shown).}\label{process_collider}
\end{figure*}  

\subsection{Cross-correlation}

For obtaining the cross-correlation between the output signals, we also need the calculation of the correlation $\langle \delta \hat{I}_T  \delta \hat{I}^{\rm in}_{A/B} \rangle$ 
where $\hat{I}^{\rm in}_{A/B}$ are the current operators before the collider. This is readily done by examining how current fluctuations in the incoming beams are transferred to the tunnel current $\hat{I}_T$. The linear form of Eq.~\eqref{tunneling-current} justifies the linear expansion
\begin{equation}
 \delta \hat{I}_T =  \delta \hat{I}_T^0 + \frac{\partial I_T}{\partial I_A} \delta \hat{I}^{\rm in}_{A}+ \frac{\partial I_T}{\partial I_B} \delta \hat{I}^{\rm in}_{B}
\end{equation}
for current fluctuations. $\delta \hat{I}_T^0$ is a notation for the part of the tunnel current fluctuations independent of the input beam fluctuations and $I_T = \langle  \hat{I}_T \rangle$.  Using that the beams generated by the fractional emitters are poissonian and independent from each other, $\langle \delta \hat{I}^{\rm in}_{A/B}  \delta \hat{I}^{\rm in}_{A/B} \rangle = e^* I_{A/B}$, we obtain 
\begin{equation}\label{noise3}
\langle \delta \hat{I}_T  \left( \delta \hat{I}^{\rm in}_{A} - \delta \hat{I}^{\rm in}_{B} \right) \rangle =  e^* \sum_{\pm} \frac{\partial I_T}{\partial I_{\pm}} I_{\mp}  =  e^* T_P I_+
\end{equation}
with the notation $I_{\pm} = I_A \pm I_B$.
We have all pieces now to compose the output cross-correlations. It is given by 
\begin{equation}\label{total-noise}
\langle \delta \hat{I}_A \delta \hat{I}_B \rangle
= - \langle \delta \hat{I}_T \delta \hat{I}_T \rangle +  \langle \delta \hat{I}_T  \delta \hat{I}^{\rm in}_{A} \rangle - \langle \delta \hat{I}_T  \delta \hat{I}^{\rm in}_{B} \rangle 
\end{equation}
with Eqs.~\eqref{noise2} and~\eqref{noise3}. The expectations values are taken at zero frequency such that the ordering of operators does not matter. 

Comparing Eqs.~\eqref{noise2} and~\eqref{noise3}, we see that the partition noise logarithmically dominates the reflected shot noise and the total cross-correlation noise is negative.
As discussed above, it implies a production of noise caused by the scattering of electrons in a genuinely non-interacting out-of-equilibrium state. The cross-correlation can be written in a normalized form, following Ref.\cite{Bartolomei2020,rosenow2016}
\begin{equation} \label{noise4}
\frac{\langle \delta \hat{I}_A \delta \hat{I}_B \rangle}{e^* I_{+} \partial_{I_{-}}  I_T } \simeq 1  -  \frac{2 \sin^2 \pi \nu}{(\pi \nu)^2}\,
\ln\left(\frac{e^* e V/h }{ 2|\sin \pi \nu| \tilde{I}(\nu) }\right). 
\end{equation}
\cm{This expression has been obtained in the regime $e^2 V / h \gg I_{A/B}$, the argument of the log is therefore large and these fluctuations are always negative.} 
We analyze hereafter this result in the two cases of $I_B=0$, where fractional charges are solely injected from the emitter $A$, and the balanced case $I_A = I_B$.

\subsection{Cross-correlation at $I_A \ne 0$ and $I_B = 0$} \label{section_onesideinjection}

The case $I_B = 0$, where Eq.~\eqref{noise4} reduces to
\begin{equation} \label{noise_A}
\frac{\langle \delta \hat{I}_A \delta \hat{I}_B \rangle}{e^* I_A \partial_{I_A}  I_T } \simeq 1  -  \frac{2 \sin^2 \pi \nu}{(\pi \nu)^2}\,
\ln\left(\frac{e^* e V/h }{ 2 I_A |\sin \pi \nu| }\right),
\end{equation}
is an insightful limit to consider. It excludes collisions between fractional charges as all charges originate from the same emitter $A$. We illustrate the underlying partition processes at the QPC in the weak backscattering limit $r \ll 1$ where fractional charges are injected very rarely from the emitter.  A fractional charge is a composition of many particle-hole pair excitations, hence, its injection triggers electron tunneling at the collider QPC from the channel A to B with rate $W_{A \to B}$ and from B to A with rate $W_{B \to A}$. They relate to the current $I_T = e (W_{A \to B} - W_{B \to A})$  and noise $\langle \delta \hat{I}_T \delta \hat{I}_T \rangle = e^2 (W_{A \to B} + W_{B \to A})$, and they are both proportional to $r^2$. The tunneling from A to B leads to a partition of the fractional charge $e^* = \nu e$ into a charge $e^* - e$ on the channel A and another $e$ on B, while the tunneling from B to A results in another partition into $e^* + e$ on A and $-e$ on B, yielding the cross-correlation
\begin{equation} \label{part-noise}
\langle \delta \hat{I}_A \delta \hat{I}_B \rangle = - e^2 (1-\nu) W_{A \to B}  - e^2 (\nu  + 1) W_{B \to A} + O(r^4).
\end{equation}
up to leading order in $r^2$. This expression, combined with Eqs.~\eqref{tunneling-current} and \eqref{noise3}, agrees with Eq.~\eqref{total-noise}.

The tunneling with rate $W_{A \to B}$ can be further decomposed into separate processes. In the ``Conventional partition", shown in Fig.~\ref{process_collider}(a), the $A$ to $B$ electron tunneling occurs simultaneously with the arrival of the fractional charge at the QPC. Instead, the new processes $a_1$ and $a_2$, shown in Fig.~\ref{process_collider}(b), correspond respectively to an early or delayed arrival of the fractional charge with respect to the electron tunneling. The two processes $a_1$ and $a_2$ add coherently, yielding the combination $(1- e^{-2i \pi \nu})$ appearing in Eqs.~\eqref{twopoint3}. Their interference $a_2^* a_1$, shown in Fig.~\ref{process_collider}(b), represents the double exchange between an electron and a fractional charge with the braiding phase $e^{-2i \pi \nu}$~\cite{Note4}. The electron tunneling in the process $a_1$ (resp. $a_2$) occurs at time $t$ (resp. time $0$). Their time separation $t$ must be longer than the temporal width $h/e V$ of a fractional charge if one wants the double exchange process $a_2^* a_1$ to involve a full fractional charge. Appendix~\ref{Appendix_detailed_calculation} 
provides the identification between these physical processes and the evaluation of the fermionic time correlator.

In the general case of $\nu <1$, the conventional partition gives a subdominant contribution to the cross-correlation noise $\propto T_P I_A$ (see Eq.~\eqref{tunneling_noise_subdomiant}) with respect to the new processes $a_{1/2}$. The latter corresponds to keeping $t_1 \simeq t_2$ between $0$ and $t$ in the evaluation of the fermionic time correlator of Eq.~\eqref{functionK} as done in Sec.\ref{sec-collider2}. 
Therefore, the double exchange process between an electron and the fractional charges of the Poissonian distribution in Eq.~\eqref{poissonian_braiding} results in the cross-correlation noise given in Eq.~\eqref{noise_A}, in which the conventional partition has been neglected. We note that the evaluation of the transferred current $I_T$, see Eq.~\eqref{tunneling-current}, involves on the contrary short time separation of $t \simeq 0$ (see Appendix~\ref{Appendix_detailed_calculation}).  

In the non-interacting case $\nu = 1$, the free electron result is recovered from the conventional partition alone of Fig.~\ref{process_collider}(a). In this case, $1 - \nu =0$ and $W_{B \to A} = 0$ in Eq.~\eqref{part-noise}, hence the first non-vanishing term in $\langle \delta \hat{I}_A \delta \hat{I}_B \rangle$ is of order $r^4$,
indicating that the cross-correlations are, even in this unbalanced limit, more negative for $\nu <1$ than for free electrons. 
The interference term $a_2^* a_1$ in Fig.~\ref{process_collider}(b) does not contribute to the cross correlations $\langle \delta \hat{I}_A \delta \hat{I}_B \rangle$ when $\nu=1$.
For free electrons, the interference term  $a_2^* a_1$ can be seen as a disconnected Feynmann diagram (a vacuum bubble diagram), since the double exchange phase factor $e^{-2 i \pi \nu} = 1$ becomes trivial, and the disconnected diagram is canceled by another ``partner" disconnected diagram (not shown) according to the linked cluster theorem.
The violation of this cancellation happens in the $\nu <1$ case, explaining why the two factors $1$ and $e^{-2 i \pi \nu}$ appear pairwise with the combination $(1- e^{-2i \pi \nu})$~\footnote{The diagram $a_2^* a_1$ in Fig.~\ref{process_collider}(b) comes with a partner diagram with opposite sign in the free fermion case. In the interacting case $\nu <1$, the two same diagrams appear, but the  $a_2^* a_1$ term comes with a $e^{-2 i \pi \nu}$ braiding phase while the partner diagram still has a factor $1$ as it does not involve braiding of particles, resulting in the overall factor $1- e^{-2i \pi \nu}$.} in Eqs.~\eqref{twopoint2} and \eqref{twopoint3}.
The partial cancellation also occurs in fractional quantum Hall systems with Laughlin anyons~\cite{han2016,lee2019}, and the fractional quantum Hall analogue of the perturbative and unbalanced limit in Fig.~\ref{process_collider} was studied in Ref.~\cite{lee2019}. 

\subsection{Cross-correlation at $I_A = I_B$}

In the balanced case $I_A = I_B$, fractional charges are injected from both emitters $A$ and $B$, and the cross-correlation takes the form 
\begin{equation} \label{noise_AB}
\frac{\langle \delta \hat{I}_A \delta \hat{I}_B \rangle}{e^* I_{+} \partial_{I_{-}}  I_T } \simeq 1  -  \frac{2 \sin^2 \pi \nu}{(\pi \nu)^2}\,
\ln\left(\frac{e^* e V/h }{4 I_A \sin^2 \pi \nu}\right).
\end{equation}
As indicated by its similarity with Eq.~\eqref{noise_A}, the cross-correlation is also determined in the balanced case by the double exchange processes shown in Fig.~\ref{process_collider}(b). Although we use the word collider for the beam splitter QPC in Fig.~\ref{fig5}, the cross-correlation in Eq.~\eqref{noise_AB} is not dominated by collisions between fractional charges arriving simultaneously at the QPC and therefore does not measure a deviation from the fermionic anti-bunching effect. Collisions occur at a rate of the order of $r^4$ whereas the cross-correlation in Eq.~\eqref{noise2}, yielding Eq.~\eqref{noise_AB}, is of order $r^2$, thereby emphasizing the prominent role of electron-anyon braiding over collisions.

Interestingly, the above analysis extends to the purely anyonic case theoretically proposed by Rosenow, Levkivskyi, and Halperin~\cite{rosenow2016} and experimentally investigated by Bartolomei {\it et al.}~\cite{Bartolomei2020}. We revisit the physical interpretation of their findings and claim that they should not be interpreted in terms of deviation from fermionic antibunching in the collision of anyons but rather as a direct measurement of the double exchange (braiding) between edge anyons, a manifestation of anyonic fractional statistics. The correlator studied in Ref.~\cite{rosenow2016}, corresponding to Eq.~\eqref{twopoint3}, is also for a long-time process with $t \gg h/ (e^* V)$, and involves a double exchange between an anyon excited at the collider QPC and the anyons of a Poisson distribution injected from a voltage-biased edge channel via anyon tunneling through a QPC.
\cm{In contrast to that, deviations from fermionic antibunching in collisions occur within the short time scale $t < h/ (e^* V)$ and do not enter this correlator~\eqref{twopoint3}. $h/ (e^* V)$ represents the time uncertainty for a direct tunneling at the collider QPC, as we would have in a collision, of an anyon injected at bias voltage $V$.}

\HS{Both the experiment by Bartolomei {\it et al.} and our work have \cm{in common} the two conditions of $k_B T \ll  e^* V$ and $I_{A/B} h/e^* \ll e^* V$ for the braiding or the double exchange. Under the condition of $k_B T \ll  e^* V$, the long-time braiding process \cm{at} $t \gg h/ (e^* V)$ results in the dominant contribution to the cross correlation, while the short-time (anti)bunching process of $t < h/ (e^* V)$ gives only a \cm{subleading} contribution. And, at $I_{A/B} h/e^* \ll e^* V$, the spatial width of the anyons injected by the voltage $V$ is much shorter than the mean distance between the anyons so that the braiding phase is well defined. 
}

Finally, the results in Eq.~\eqref{noise2} and Eq.~\eqref{noise_AB}, up to the logarithmic factor, depend universally on the exchange phase $\pi \nu$, similarly to the purely fractional case~\cite{rosenow2016,Bartolomei2020} albeit with a different functional form. The difference originates from the field dimension at the collider QPC, involving electron in the present case as opposed to anyons in the purely fractional situation of Refs.~\cite{rosenow2016,Bartolomei2020}. We can view the excitation induced by our emitter as an hybrid object composed by an equilibrium fermionic side married to a non-equilibrium anyonic part.

\section{Summary and outlook}\label{sec-summary}

We proposed a many-terminal geometry in the integer quantum Hall effect that realizes an emitter of fractional charges. The set-up comprises $N$ ballistic channels covered in the central region by a metallic island imposing charge conservation. An additional $N+1$ channel is tuned in the quasi-ballistic  regime with a quantum point contact close to the island. As electrons are excited towards the island, they scatter and separate into fractional charges towards the output channels. Only continuous bias are used, {\it i.e.} the set-up is not operated with time-dependent pulses. As the quantum point contact is only weakly reflective, the train of fractional charges is dilute and exhibits poissonian statistics  as we proved by computing the full counting statistics.

We further characterized these beams of fractional charges by investigating their mixing through a quantum point contact acting as a beam splitter. By measuring the current noise cross-correlations of the output signals, one finds a negative correlation. It differs from the same calculation for free fermions where such correlations are absent, and indicate that the incoming fractional beam incorporates an out-of-equilibrium collection of electron-hole excitations. The cross-correlations  depend essentially on the mutual statistical phase between the electrons tunneling at the collider QPC and the fractional charges emulating anyons sent by the emitters. Coincidence processes where fractional charges arrive simultaneously at the collider and exhibit imperfect anti-bunching were identified as being higher order in the beam currents and discarded. The braid or exchange process of anyons also prevails over the bunching physics in the setup of Refs.~\cite{rosenow2016,Bartolomei2020}.

\begin{figure}
\centering\includegraphics[width=0.9\columnwidth]{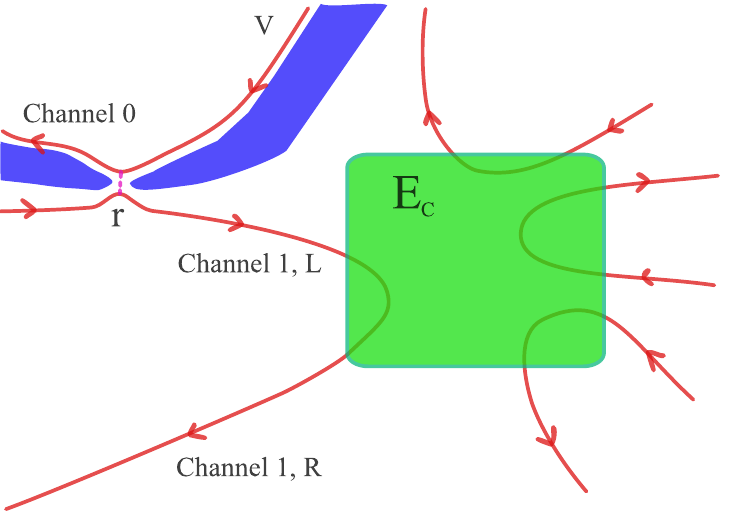}
\caption{\cm{Alternative device for the fractional emitter where the first partitioner quantum point contact is located upstream from the island. The voltage $V$ is applied on channel $0$, {\it i.e.} not directly connected to the island, thereby strongly reducing Joule heating. Electrons in channel $0$ are reflected into channel \HS{$1,L$}, and thus enter the island, with the amplitude $r \ll 1$. The output channel is $1,R$ \HS{and} also denoted $A$ with the field $\psi_A$.}  \label{fig7}}
\end{figure}

\cm{In a realistic implementation of the proposal shown in Fig.~\ref{fig4}, Joule heating plays an important and detrimental role against the measurement of fractional braiding. Indeed, for a charging energy on the order of \HS{300 mK} as in the experiment of Ref.~\cite{Anthore2018}, the circuit needs to be operated at low temperatures and the electron-phonon coupling becomes insufficient for evacuating heat out of the island~\cite{Sivre2019}. The power generated by the voltage sources, and transferred to the island, is on the order of $\sim V^2/R_q$ as all channels are nearly ballistic. In the absence of electron-phonon cooling, this is balanced by the heat power sent out to the outgoing chiral edges: $(\pi k_B T_\Omega)^2 /6 h$ per channel, assuming an equilibrated temperature $T_\Omega$ on the island and zero-temperature in the reservoirs~\cite{Jezouin2016,Sivre2018}. The resulting island temperature is stabilized at \HS{$T_\Omega \sim e V/(\sqrt{N} k_B)$} and thermal noise is likely to mask fractional charges in the output beam. Fortunately, there are ways out. Current experiments are developing islands with much smaller capacitances corresponding to higher charging energies. They can be studied at temperature where electron-phonon cooling is again efficient in thermalizing the island. Alternatively, the device shown in \HS{Fig.~\ref{fig7}} can be implemented, where the voltage source is no longer directly connected to the island, thereby minimizing Joule heating. Electrons leaving the voltage source and later entering the island must be reflected by the QPC in \HS{Fig.~\ref{fig7}} with the probability amplitude $r \ll 1$. Repeating the analysis of Sec.~\ref{sec-collider}~\cite{Note3}, we obtain for the one-body correlation function in the output channel:
\begin{equation}\label{twopointnew}
\langle \psi_A(t)\psi^\dagger_A(0)\rangle \simeq \frac{1}{2 \pi v_F}  \frac{ e^{-  \frac{I_A}{e} \left( 1 - e^{- 2 i \pi \nu_1}  \right)  \,  t}} {a+i t}. 
\end{equation}
with $\nu_1 = 1/(N+1)$ and $I_A = r^2 e^2 V/h$, indicating that the QPC does not suffer from dynamical Coulomb blockade by the island. Eq.~\eqref{twopointnew} corresponds to the emission of fractional charges $e^*= \nu_1 \, e$ and the result of two such sources through a third collider QPC reproduces Eq.~\eqref{noise2} with $\nu_1$ replacing $\nu$.}

Our findings open many perspectives as the collision between non-identical emitters can also be envisioned with hole and/or electron current and different fractional charges; this paper focused on identical emitters. The statistical properties of the beams originating from the collider would also be useful to detail the processes of charge transfer. Finally, we have restricted in this paper our analysis to constant voltage biases but it would be an appealing direction to analyse the collision of beams with fractional properties inherited from time-dependent excitations~\cite{Freulon2015,wahl2014}, e.g. by lorentzian voltage pulses~\cite{dubois2013,dubois2013-2}.

\begin{acknowledgments}
We would like to thank A. Anthore and F. Pierre for insightful discussions and for providing us the motivation to investigate this problem. Discussions with F. von Oppen and  G. F\`eve are also gratefully acknowledged. This work was supported by the French National Research Agency (project SIMCIRCUIT, ANR-18-CE47-0014-01) and Korea NRF (SRC Center for Quantum Coherence in Condensed Matter, 2016R1A5A1008184; NRF-2019-Global Ph.D. Fellowship Program).
\end{acknowledgments}

\appendix

\section{Rotation to new Bose fields}\label{rotation}
 We describe the rotation from the original fields $\phi$ to the new fields $\tilde{\phi}$. It takes the form of $\tilde{\phi}_i = O_{i,j} \phi_j$ where $O$ is an orthogonal matrix, {\it i.e.} $O^t O = 1$. We give here the expressions of the first three lines of $O$,
\begin{equation}
\begin{split}
\tilde{\phi}_1 &=\frac{1}{\sqrt{N+1}}\displaystyle\sum_{j=1}^{N+1}\phi_j\\
\tilde{\phi}_2 &=\sqrt{\frac{N}{N+1}}\Big(\phi_1-\frac{1}{N}\displaystyle\sum_{j=2}^{N+1}\phi_j\Big)\\
\tilde{\phi}_3 &=\sqrt{\frac{N-1}{N}}\Big(\phi_2-\frac{1}{N-1}\displaystyle\sum_{j=3}^{N+1}\phi_j\Big)
\end{split}
\end{equation}
whereas the expression of the other new fields do not play a role. With these definitions, we can expressed the first two original fields,
\begin{equation}
\begin{split}
\phi_1&=\frac{1}{\sqrt{N+1}}\,\tilde{\phi}_1+\sqrt{\frac{N}{N+1}}\,\tilde{\phi}_2\\
\phi_2&=\frac{1}{\sqrt{N+1}}\,\tilde{\phi}_1-\frac{1}{\sqrt{N(N+1)}}\tilde{\phi}_2+\sqrt{\frac{N-1}{N}}\,\tilde{\phi}_3
\end{split}
\end{equation}
in terms of the new fields $\tilde{\phi}_1$, $\tilde{\phi}_2$ and $\tilde{\phi}_3$.

\section{Circuit description}\label{circuit}

We detail here the solution to the free bosonic problem of Fig.~\ref{fig1} where all channels are ballistic. It can be described by means of a scattering theory of plasmonic waves similar to the input/ouput theory of quantum circuits. The Hamiltonian $H = H_0 + H_c$ gives Heisenberg equations of motion for the bosonic fields introduced in appendix~\ref{rotation}. In the absence of external voltages, they are expressed as
\begin{subequations}\label{EOM1}
\begin{align}
\label{EOM11}
\partial_t\tilde{\phi}_{R,j} &= -v_F\partial_x\tilde{\phi}_{R,j}-\frac{2E_c}{\hbar}\delta_{1,j}\sqrt{N+1}\,\hat{N} \theta(-x),\\
\label{EOM12}
\partial_t\tilde{\phi}_{L,j} &= v_F\partial_x\tilde{\phi}_{L,j}-\frac{2E_c}{\hbar}\delta_{1,j}\sqrt{N+1}\,\hat{N} \theta(-x),
\end{align}
\end{subequations}
where only the modes $j=1$ see the charging energy and the total charge $\hat{N}$ below the island, see Eq.~\eqref{total-charge}. The solution to Eq.~\eqref{EOM1} is given by~\cite{Slobodeniuk2013}
\begin{equation}\label{solEOM1}
\begin{split}
\tilde{\phi}_{R,1}(x,t)&=\tilde{\phi}^{(0)}_{R,1}(x,t)
-\frac{1}{\tau}\int_{-\infty}^{t-\frac{x}{v_F}}\text{d}t'\,e^{-\frac{t-t'-x/v_F}{\tau}}\tilde{\varphi}(t') \\
\tilde{\phi}_{L,1}(x,t)&=\tilde{\phi}^{(0)}_{L,1}(x,t)
\end{split}
\end{equation}
for outside the dot $x>0$, and
\begin{equation}\label{solEOM2}
\begin{split}
\tilde{\phi}_{R,1} & (x,t)=\tilde{\phi}^{(0)}_{R,1}(x,t)-\frac{1}{\tau}
\int_{-\infty}^{t}\text{d}t'e^{-\frac{t-t'}{\tau}}\tilde{\varphi}(t')\\
\tilde{\phi}_{L,1} & (x,t)=\tilde{\phi}^{(0)}_{L,1}(x,t) \\[2mm] & -\frac{1}{\tau}
\int_{-\infty}^t\text{d}t'\,e^{-\frac{t-t'}{\tau}} \left [\tilde{\varphi}(t') 
 -\tilde{\varphi}(t'+\frac{x}{v_F}) \right]
\end{split}
\end{equation}
for inside the dot $x<0$, where it is clear that the two expressions match at $x=0$. We have introduced here $\tilde{\varphi}(t')=\tilde{\phi}^{(0)}_{R,1}(0,t')-\tilde{\phi}^{(0)}_{L,1}(0,t')$ and $\tau=\pi\hbar/(N+1)E_c$. The subscript $(0)$ refers to the free solutions, {\it i.e.} the solutions of Eq.~\eqref{EOM1} in the absence of charging energy, $E_c = 0$, and therefore of scattering. They can be decomposed into propagating incoming modes (see  Eq.~\eqref{expansion})
\begin{equation}\label{expansion2}
\tilde{\phi}_{R /L,1}^{(0)} (x,t) = \int_0^{+\infty} \frac{d \omega}{\sqrt{\omega}} \left( b/a_{in,\omega} e^{-i \omega (t \mp x/v_F)} + h.c. \right). 
\end{equation}
After scattering, the outgoing solution~\eqref{solEOM1} ($x>0$) can also be decomposed as
\begin{equation}\label{expansion3}
\tilde{\phi}_{R,1} (x,t) = \int_0^{+\infty} \frac{d \omega}{\sqrt{\omega}} \left( a_{out,\omega} e^{-i \omega (t - x/v_F)} + h.c. \right), 
\end{equation}
and the same expression holds for $\tilde{\phi}_{L,1} (x,t)$, $x<0$, with $a_{out,\omega}$ replaced by $b_{out,\omega}$ and $t-x/v_F$ by $t+x/v_F$. The geometry of the different modes is sketched in Fig.\ref{fig3}. Inserting the above expansions over plane waves into Eqs.~\eqref{solEOM1} and~\eqref{solEOM2}, we obtain
\begin{equation}
    a_{out,\omega}=b_{in,\omega}-\frac{1}{1-i\tau\omega}(b_{in,\omega}-
    a_{in,\omega})
 \end{equation}
and the input/output S matrix written in Eq.~\eqref{sca1}.

The scattering problem solved here is exactly the same as the circuit geometry of Fig.~\ref{fig3} with one capacitance connecting two transmissions lines. To properly map the two models, we identify the canonically conjugated charge  and 
 voltage operators  $(\hat{Q},\hat{V})$ of the transmission lines as 
\begin{equation}
\hat{Q} (x,t) = \frac{e}{2 \pi}  \left(\tilde{\phi}_{L,1} (x,t) - \tilde{\phi}_{R,1} (x,t) \right) 
\end{equation}
and $\hat{V} (x,t)  =   (\pi \hbar v_F/e^2)  \partial_x \hat{Q} (x,t)  $ corresponding to a characteristic impedance of $R_q/2=\pi \hbar/e^2$. The bridging capacitance is $C/(N+1)$, renormalized by the total number of channels.

\section{Bosonization}\label{appen-bosonization}

We give the standard commutators and the average values of the relevant  products of bosonic fields at zero temperature.
The rotated bosonic fields $\tilde{\phi}_{\alpha,i}$  with $i=1,\ldots,N+1$  and $\alpha=L,R$ satisfy diagonal commutation rules. For $i\neq1$,
\begin{equation}
\begin{split}
[\tilde{\phi}_{\alpha,i}(t),\tilde{\phi}_{\beta,j}(t')]&=-i\pi\,\text{sgn}(t-t')\,\delta_{ij}\delta_{\alpha\beta}\\
\langle \tilde{\phi}_{\alpha,i}(t)\tilde{\phi}_{\beta,j}(t')\rangle -\langle\tilde{\phi}_{\alpha,i}^2\rangle &=\delta_{ij}\delta_{\alpha\beta}
\ln\Big(\frac{a}{a+i(t-t')}\Big)
\end{split}
\end{equation}
As a consequence, we have these  relations for the field  $h$,
\begin{equation}
\begin{split}
[h(t),h(t')]&=-i2\pi\nu\,\text{sgn}(t-t')\\
[h(t),\phi_{R,1}(t')]&=-i\pi\nu\,\text{sgn}(t-t')\\
\langle h(t)h(t')\rangle-\langle h^2\rangle &=2\nu\ln\Big(\frac{a}{a+i(t-t')}\Big)
\end{split}
\end{equation}
In  the channel 1, we have massive fields. The bosonic fields $\tilde{\phi}_{R,1}$ follow the same relations (same for 
$\tilde{\phi}_{L,1}$) but   
\begin{equation}
\begin{split}
[\tilde{\phi}_{R,1}(t),\tilde{\phi}_{L,1}(t')]&=\int_{-\infty}^{+\infty}\frac{\text{d}\omega}{\omega}\,\frac{e^{-i\omega(t-t')}}{1-i\tau\omega}\\
\langle\tilde{\phi}_{R,1}(t)\tilde{\phi}_{L,1}(t')\rangle&=\int_{0}^{+\infty}\frac{\text{d}\omega}{\omega}\,\frac{e^{-i\omega(t-t')}}{1-i\tau\omega}
\end{split}
\end{equation}
 where $\tau=\pi\hbar/(N+1)E_c$.
 Now, we give a few details on how one goes from Eq. \eqref{tunneling2} to Eq. \eqref{tunneling3}. We integrate Eq. \eqref{tunneling2} over the massive mode $\tilde{\phi}_{L/R,1}$. We need to compute
 
 \begin{equation}
     \langle e^{\frac{i}{\sqrt{N+1}}(\tilde{\phi}_{L,1}-
     \tilde{\phi}_{R,1})}\rangle = e^{-\frac{1}{2(N+1)}
     \langle(\tilde{\phi}_{L,1}-
     \tilde{\phi}_{R,1})^2\rangle}
      \end{equation}
Using the relations in  this appendix, we can perform the average value and find
 
\begin{multline}
\langle(\tilde{\phi}_{L,1}-
     \tilde{\phi}_{R,1})^2\rangle=2\int_0^{+\infty}
     \text{d}\omega\,\frac{\tau^2\omega}{1+\tau^2\omega^2}e^{-\omega\hbar/D}\\[3mm]
     \simeq 2\ln\Big(\frac{\pi D}{e^\gamma(N+1)E_c}\Big)
\end{multline}
 Finally, 
  
   \begin{equation}
     \langle e^{\frac{i}{\sqrt{N+1}}(\tilde{\phi}_{L,1}-
     \tilde{\phi}_{R,1})}\rangle =
     \Big(\frac{e^\gamma (N+1)E_c}{\pi D}\Big)^{1/(N+1)}
     \end{equation}
      
\section{Alternative output channel}\label{alternative}

The focus in most of this work is on the use of channel $1$ to emit fractional charges, see Fig.\ref{fig1}. Alternatively, one can use another channel. Without any loss of generality, we consider the channel $2$ in Fig.\ref{fig1} as the output. The current operator for the outgoing edge state is 
$\hat{I}_{R,2}=-(e/2\pi)\partial_t\phi_{R,2}$ or, using Appendix~\ref{rotation},
\begin{equation}
  \hat{I}_{R,2}=-\frac{e}{2\pi}\partial_t\Big(\frac{1}{\sqrt{N+1}}\tilde{\phi}_{R,1}-\frac{1}{\sqrt{N(N+1)}}\tilde{\phi}_{R,2}\Big)  
\end{equation}
In  this expression, the field $\tilde{\phi}_{R,3}$ has been discarded as it decouples from all other fields in the Hamiltonian and has a vanishing expectation value. Removing also the massive field $\tilde{\phi}_{R,1}$ leads to
\begin{equation}
  \hat{I}_{R,2}=\frac{e}{2\pi}\frac{1}{\sqrt{N(N+1)}}\partial_t\tilde{\phi}_{R,2}.
  \end{equation}
Starting from this expression and following the reasoning of Sec.\ref{sec-emitter}, we recover the current operators of Eqs.~\eqref{currents} with the new effective charge $e^* = \nu_1 e$ and
\begin{equation}
\nu_1 = - \frac{1}{N+1},    
\end{equation}
corresponding to a train of fractional charges with the Fano factor $F = -e/(N+1)$. However, the operator ${\cal T}$ still has dimension $\nu$ and the current is again given by Eq.~\eqref{current} (we take $V<0$ for electron instead of hole emission) but with $e^* = -e/(N+1)$ whereas $\nu= N/(N+1)$.

If we now turn to the collision of two such fractional charges $e^* = \nu_1 e$, the field correlator takes a similar form 
\begin{equation}\label{twopoint-new}
\langle \psi_A(t)\psi^\dagger_A(0)\rangle= \frac{1}{2 \pi v_F}  \frac{ e^{-  \frac{I_A}{e^*} \left( 1 - e^{- 2 i \pi \nu_1}  \right)  \,  t}} {a+i t}, 
\end{equation}
in terms of the effective charge $\nu_1$ and the incoming current $I_A$ given by Eq.~\eqref{current}. We stress again that $I_A$ depend on $\nu = N/(N+1)$ as well as on the charge $e^* = \nu_1 e$. However, once the cross-correlations are written in terms of the fractional charge $\nu_1$ and the currents $I_{A/B}$, we find precisely the same expressions as in the main text with the replacement of $\nu$ by $\nu_1$.

\section{Counting fields}\label{counting-field}

The evaluation of the whole distribution of transmitted charges of Sec.~\ref{sec-FCS} relies on the introduction of counting fields which probe the flow of electrons. This is done by adding a source term to the Hamiltonian  $H_\lambda = H +\hat{V}_\lambda$, where $\hat{V}_\lambda= - \hbar \lambda \hat{I}_{R,1} /(2e)$ couples the counting field $\lambda$ to the outgoing current $\hat{I}_{R,1}$ taken at a distance $L$ from the island. The first Heisenberg equation of motion~\eqref{EOM11} derived in appendix~\ref{circuit} acquires the additional source term 
\begin{equation}
\frac{i}{\hbar}[\hat{V}_\lambda, \tilde{\phi}_{R,j}]=\frac{\lambda v_F}{2\sqrt{N+1}}\delta(x-L)\,(\delta_{1,j}+\sqrt{N}\delta_{2,j})
\end{equation}
localized at $x=L$, whereas the second one~\eqref{EOM12} is not modified. The solution with this source term proceeds additively as compared to appendix~\ref{circuit}. One obtains the expressions of Eqs.~\eqref{solEOM1} and~\eqref{solEOM2} shifted by
\begin{equation}
\begin{split}
\tilde{\phi}_{R,1} (x,t) &\to \tilde{\phi}_{R,1} (x,t) - \frac{\lambda}{2\sqrt{N+1}}
\theta(L - x) \\
\tilde{\phi}_{L,1} (x,t) &\to \tilde{\phi}_{L,1} (x,t) - \frac{\lambda}{2\sqrt{N+1}} \\
\tilde{\phi}_{R,2} (x,t) &\to \tilde{\phi}_{R,2} (x,t) - \frac{\lambda}{2\sqrt{N+1}}
\theta(L - x)
\end{split}
\end{equation}
For the total charge mode $j=1$, the shifts of the $R$ and $L$ fields leave the charge $\hat{N}$ invariant. For $j=2$ on the contrary, the $R$ and $L$ fields are not coupled and only the $R$ field is shifted.

With this modified solution, the tunneling operator $\cal{T}$ is dressed as
\begin{equation}
\mathcal{T}(t) \to e^{i \nu\lambda/2}\,\mathcal{T}(t),
\end{equation}
indicating that $\lambda$ measures a charge $\nu$ for each tunneling event as represented in Fig.~\ref{fig4}. Note that the shifts for $j=1$ cancel each other. If, alternatively, one probes the charge emitted in the ballistic channel $2$ via the coupling $\hat{V}_\lambda= - \hbar \lambda \hat{I}_{R,2} /(2e)$, then $\tilde{\phi}_{R,2} (x,t)$ now shifts by $+ \lambda/ 2 \sqrt{N(N+1)}
\theta(L - x)$ and the tunneling operator is dressed as $e^{-i \nu_1\lambda/2}\,\mathcal{T}(t)$ (see appendix~\ref{alternative}), corresponding to hole tunneling with fractional charge $\nu_1$.

Finally, the generating function~\eqref{fcs1} can be expanded to second order in $r$, the first order being zero, where one needs to evaluate the following term
\begin{equation}\label{interme1}
   \sum_{\eta_{1,2}} \eta_1\eta_2
    \int_0^t dt_{1}\int_0^t  dt_{2} e^{i(\eta_1-\eta_2)\frac{\lambda\nu}{2}} \langle\mathcal{T}(t_1)\mathcal{T}^\dagger(t_2)\rangle
\end{equation}
The times $t_{1/2}$  are  taken with  respect to the contour index $\eta_{1,2}$. Changing the time integration to the center of mass $(t_1+t_2)/2$ and relative times $\tau = t_1 - t_2$, one approximates Eq.~\eqref{interme1} for long time $t$ as ($t>0$)
\begin{equation}
\left(\frac{{\cal N}}{a^{-\nu}} \right)^2 t    \sum_{\eta_{1,2}} \eta_1\eta_2 \int_{-\infty}^{+\infty} d \tau  \,  \frac{e^{-i e V \tau/\hbar}e^{i \pi \nu (\eta_1 - \eta_2)}}{[a+i \tau\,\chi_{1,2}(\tau)]^{2 \nu}}  
\end{equation}
where $\chi_{1,2}(\tau)$ is defined just after Eq.~\eqref{functionK}. The sum of the $\eta_1,\eta_2 = ++$ and $--$ Keldysh terms equals the sum of the $+-$ and $-+$ terms (causality). Checking that the $-+$ component vanishes for $V>0$, we eventually obtain Eq.~\eqref{charac2}.

\section{Tunneling current and noise}\label{Appendix_detailed_calculation}

Here we provide detailed calculations of the tunneling current and noise at the collider QPC in Eqs.~\eqref{ITcurrent}  and \eqref{noise}. The correlator in Eq.~\eqref{twopoint} is computed as
\begin{equation} \label{shorttime}
\langle \psi_A(t)\psi^\dagger_A(0)\rangle=  \frac{1}{2\pi v_F}\frac{1}{a+it}(1-i \frac{2\pi I_A}{e} t + \mathcal{O}(t^2))
\end{equation}
for $|t| \lesssim h/eV$, and
\begin{widetext}
\begin{equation} \label{longtime}
\langle \psi_A(t)\psi^\dagger_A(0)\rangle \simeq \frac{1}{2 \pi v_F}   \frac{1}{ a + it}  \big[ e^{-  \frac{I_A}{e^*} \left( 1 - e^{- 2 i \pi \nu \text{sgn}(t)}  \right)  \,  |t|} + \frac{I_A h}{e^2 V} \frac{\Gamma (2 \nu)}{\nu \Gamma (\nu)^2} (e^{-i e V t/\hbar} - 1 + e^{-i \pi \nu \textrm{sgn} (t)} - \cos (\pi \nu) ) + \cdots \big]
\end{equation}
\end{widetext}
for $|t| \gg h/eV$. The first term of Eq.~\eqref{longtime} is discussed in the main text, and obtained from the integral time  domain of $0 \ll t_1 \simeq t_2 \ll t$ and $t \gg h / (eV)$ in Eq.~\eqref{functionK}. It corresponds to the two processes $a_1$ and $a_2$ shown in Fig.~\ref{process_collider}(b), providing the dominant contribution. 
The other terms, called conventional partition and shown Fig.~\ref{process_collider}(a) in the main text, are subdominant and neglected in Eq.~\eqref{twopoint3}, and they are obtained from the integral time domains of $t_1 \simeq 0$ and $t_2 \simeq t$,  $t_1 \simeq t_2 \simeq 0$, or $t_1 \simeq t_2 \simeq t$. As $\nu \to 1$, Eqs.~\eqref{shorttime} and \eqref{longtime} become identical to those of the free electron case.

We compute the tunneling current and noise at the collider QPC in Eqs.~\eqref{ITcurrent}  and \eqref{noise},
with applying Eq.~\eqref{shorttime} to the integral range of $|t| < h / (\alpha eV)$ and 
Eq.~\eqref{longtime} to $|t| > h / (\alpha eV)$. Here $\alpha$ is of the order of 1.
The results are independent of $\alpha$; this justifies the cutoff replacement of $a \to h/(eV)$ in the calculation of the dominant contribution to $\langle \delta \hat{I}_T \delta \hat{I}_T \rangle$. For the tunneling current $I_T$, the small $t$ behavior is dominant and we reproduce the result of Eq.~\eqref{tunneling-current} obtained in the main text from a straightforward Landauer-B\"utticker analysis. We obtain the noise as
\begin{equation} \label{tunneling_noise_subdomiant}
\frac{\langle \delta \hat{I}_T \delta \hat{I}_T \rangle}{e^2 T_P} = \frac{2 I_+ \sin^2 \pi \nu}{\pi^2 e^*} \ln\left(\frac{e^* e V/h }{2|\sin \pi \nu| \tilde{I} (\nu) }\right) + \frac{I_+ \Gamma (2 \nu)}{e^* \Gamma (\nu)^2}  +  \cdots,
\end{equation}
where $\tilde{I} (\nu) = \sqrt{I_+^2 \sin^2 \pi \nu + I_-^2 \cos^2 \pi \nu}$ and $I_\pm = I_A \pm I_B$.
The first logarithmic term originates from the partition processes with the double exchange in Fig.~\ref{process_collider}(b), while the second term linear in $I_+$ comes from the conventional partition process in Fig.~\ref{process_collider}(a).
When $\nu = 1$, the free electron results of $\langle \delta \hat{I}_T \delta \hat{I}_T \rangle = e T_P I_+$ (up to the linear order in $T_p$) and $\langle \delta I_A \delta I_B \rangle = 0$ for $I_A = I_B$ are reproduced.

We emphasize that even in the case of $I_A = I_B$ where fractional charges are emitted from both the emitters A and B, the two terms originate from the partition processes in Figs.~\ref{process_collider}(a) and (b). The collision process, which happens when a fractional charge on the channel A and another on the channel B arrive simultaneously at the collider QPC, provides a much weaker contribution to the cross correlation and are not shown in the above equation.

\bibliography{biblio.bib}

\clearpage

\setcounter{figure}{0}
\setcounter{section}{0}
\setcounter{equation}{0}
\renewcommand{\theequation}{S\arabic{equation}}
\renewcommand{\thefigure}{S\arabic{figure}}

\newcommand{\paf}[2]{\left(\frac{#1}{#2}\right)}
\newcommand{\ex}[1]{\mathrm{e}^{\,#1}\xspace}

\onecolumngrid

\section*{\Large{Supplemental material}}

\section{Output field correlation function}
\subsection{Second order expansion}
We are interested in the evaluation of the fermionic correlator
\begin{equation}\label{twopoint-sm}
\langle \psi_A(L,t) \psi^\dagger_A(L,0)\rangle= \langle T_K \psi_A(L,t^-)\psi^\dagger_A(L,0^+)
e^{-i \int_K d t' H_{BS} (t') /\hbar} \rangle 
\end{equation}
for the output channel of the fractional emitter $A$ (see the notation introduced in Sec. IV of the main text). $\psi_A$ stands for the output field $\psi_{R,1}$.
For completeness, we recall the expression of the tunneling Hamiltonian $H_{BS} = \hbar v_F r \left( {\cal T} + {\cal T}^\dagger \right)$ and the bosonization formulas for the field and tunnel operators
\begin{equation}\label{bosonization}
\psi_{A} (x) = { \sqrt{\dfrac{D}{2\pi \hbar v_F}} e^{i\phi_{A}(x)}} \qquad {\cal T} (t) = {\cal N} e^{i\sqrt{\frac{N}{N+1}} \, ( \tilde{\phi}_{L,2} - \tilde{\phi}_{R,2} )} \, e^{- i e V t/\hbar}
\end{equation}
where 
\begin{equation}
{\cal N} = \frac{D}{2 \pi \hbar v_F}\left(\frac{e^\gamma (N+1)E_c}{\pi D}\right)^{1/(N+1)}, \qquad 
\phi_{A} = \frac{1}{\sqrt{N+1}} \, \tilde{\phi}_{R,1} + \sqrt{\frac{N}{N+1}} \, \tilde{\phi}_{R,2}, 
\end{equation}the bosonic field $\phi_A$ has been expressed in terms of the rotated bosonic modes.  Expanding Eq.~\eqref{twopoint-sm} to second order in $r \ll 1$, we perform the average over the quadratic bosonized Hamiltonian and obtain
\begin{equation}
\langle \psi_A(L,t)\psi^\dagger_A(L,0)\rangle= \frac{1}{2 \pi v_F}  \frac{1}{a+i  t}  \left[ 1 - \left(\frac{v_F r {\cal N}}{a^{-\nu}} \right)^2 {\cal K} (t)\right], 
\end{equation}
with the short-time cutoff $a=\hbar/D$ and
\begin{equation}\label{functionK}
\begin{split}
{\cal K} (L,t) = \sum_{\eta_{1/2}} \eta_1\eta_2
\int_{-\infty}^{+\infty} d t_1\,\int_{-\infty}^{+\infty} d t_2
\, e^{i e V (t_1-t_2)/\hbar}\,\paf{1}{a+i (t_1-t_2)\,\chi_{\eta_1\eta_2}(t_1-t_2)}^{2 \nu}  \\  
 \times  \paf{a+i (t-t_1-L/v_F)\,\chi_{-\eta_1}(t-t_1)}{a+i (t-t_2-L/v_F)\,\chi_{-\eta_2}(t-t_2)}^{\nu}
 \paf{a-i (t_2+L/v_F)\,\chi_{+\eta_2}(-t_2)}{a-i (t_1+L/v_F)\,\chi_{+\eta_1}(-t_1)}^{\nu}
  \end{split}
 \end{equation}
 We use the notation $\chi_{\eta_1\eta_2}(t_1-t_2) = \{ {\rm sgn} (t_1-t_2) 
 (\eta_1 + \eta_2 ) -   (\eta_1 - \eta_2 ) \}/2 $. We probe the correlation far from the emitter so that $L$ is large and the regions where $t-t_{1/2}$ and $-t_{1/2}$ are close to $L/v_F$ are dominant. Hence, we can replace 
 $\chi_{-\eta_1}(t-t_1) = \eta_1$, $\chi_{-\eta_2}(t-t_2) = \eta_2$, $\chi_{+\eta_1}(-t_1) = \eta_1$ and $\chi_{+\eta_2}(-t_2) = \eta_2$. After these simplifications, $L/v_F$ can be absorbed by a shift in the time integrals and the second-order correction takes the form
 \begin{equation}\label{expreK}
 {\cal K} (t) = \sum_{\eta_{1/2}} \eta_1\eta_2
\int_{-\infty}^{+\infty} d t_1\,\int_{-\infty}^{+\infty} d t_2 \, \frac{e^{i e V (t_1-t_2)/\hbar}}{[a+i (t_1-t_2)\,\chi_{\eta_1\eta_2}(t_1-t_2)]^{2 \nu}} \\[2mm]
 \left(\frac{a- i t_2 \,\eta_2}{a-i t_1 \,\eta_1} \right)^{\nu}
\left(\frac{a+ i  (t-t_1)\,\eta_1}{a+i (t-t_2)\,\eta_2} \right)^{\nu}.
  \end{equation}
 Causality is responsible for the following identity 
\begin{equation}\label{causality}
\sum_{\eta_{1/2}}^{}\eta_1\eta_2
\paf{1}{a+i (t_1-t_2)\,\chi_{\eta_1\eta_2}(t_1-t_2)}^{2 \nu}  = 0
 \end{equation}
which can be explicitly checked regardless of the values of $t_1$ and $t_2$. Without loss of generality, we assume $t>0$ and focus on the long-time asymptotic behavior. Using causality, we find that the time integration can be restricted to the domain where $t_1$ and $t_2$ are taken between $0$ and $t$. Moreover, the integral~\eqref{expreK} is dominated by the region $t_1 \simeq t_2$.  Changing variables to the center of mass and relative time, denoted $\tau = t_1 - t_2$, one finds in the limit of large time $t$
  \begin{equation}
 {\cal K} (t) = t \sum_{\eta_{1/2}}\eta_1\eta_2 \int_{-\infty}^{+\infty} d \tau  \,  \frac{e^{i e V \tau/\hbar}}{[a+i \tau\,\chi_{\eta_1\eta_2}(\tau)]^{2 \nu}}  \, e^{i \pi \nu (\eta_1 - \eta_2)}
 \end{equation}
Performing the integral, 
\begin{equation}\label{integral}
\int_{-\infty}^{+\infty} d \tau  \,  \frac{e^{i e V \tau/\hbar}}{(a+i \tau)^{2 \nu}}  = 2 \sin( \pi \nu) \Gamma(1 - 2 \nu)  (e V/\hbar)^{2 \nu-1},
\end{equation}
for $V>0$, and the integral vanishes identically for $V<0$, we obtain
  \begin{equation}
 {\cal K} (t) =  \left( 1 - e^{- 2 i \pi \nu}  \right) 2 \sin( \pi \nu) \Gamma(1 - 2 \nu)  (e V/\hbar)^{2 \nu-1}  t 
 \end{equation}
 for $V>0$. The final result can be expressed using the output current,
 \begin{equation}\label{current-ms}
I_A = e^* r^2 \frac{e V}{\hbar} \left( \frac{\tilde{E}_c}{e^* V} \right)^{2 (1-\nu)} \frac{ \Gamma(1-2 \nu) \sin 2 \pi \nu}{2 \pi^2},
\end{equation}
which yields 
\begin{equation}\label{twopoint2}
\langle \psi_A(L,t)\psi^\dagger_A(L,0)\rangle= \frac{1}{2 \pi}  \frac{1}{a+i t}  \left[ 1 -  \frac{I_A}{e^*} \left( 1 - e^{- 2 i \pi \nu}  \right)  \,  t \right],
\end{equation}
already advertised in the main text. The long-time behavior is not captured by this expression, since the perturbative correction eventually outweighs the unperturbed result at large $t$.  

\subsection{Non-perturbative resummation}

The above perturbative approach is limited to intermediate times and a non-perturbative analysis is necessary to explore the long-time limit. Nonetheless, essentially the same computation can be extended to arbitrary order in $r$ as we will now show. 
Before proceeding, it is instructive to go again over the main features of the above leading order calculation. The main contribution is obtained for tunneling times $t_1$, $t_2$ taken close to each other and between $0$ and $t$. Integrating the product of tunneling operators ${\cal T}$ over the time difference $t_2-t_1$ gives a result proportional to the output current $I_A$. In addition, the $1 - e^{- 2 i \pi \nu} $ prefactor originates from the Keldysh ordering of the tunneling operator ${\cal T}^\dagger$ with respect to the fields $\psi_A (L,t)$ and $\psi_A^\dagger (L,0)$. ${\cal T}^\dagger$ is ordered right to the field operators when the Keldysh indices $\eta_{1/2}$ are both $+1$, whereas it moves left to them as $\eta_1=-1$, $\eta_2=+1$, thereby inducing two commutation relations
\begin{equation}
{\cal T}^\dagger (t_1) \psi_A(L,t)\psi^\dagger_A(L,0) = \psi_A(L,t)\psi^\dagger_A(L,0)  {\cal T}^\dagger (t_1)
e^{- 2 i \pi \nu} \end{equation}
since $0< t_1 < t$.
The term with $\eta_1=+1$, $\eta_2=-1$ vanishes by causality because the corresponding integral has a pole in the wrong complex half-plane. Apart from this double commutation, or braid operation, the computation of the two resulting contributions are the same with the Keldysh orientation factor $\eta_1 \eta_2=+1$ in one case and $\eta_1 \eta_2=-1$ in the other case, such that one obtains an overall $ 1 - e^{- 2 i \pi \nu} $ factor. 

Now, let us consider the order $2 n$ perturbative correction to the fermionic correlator (odd orders vanish)
\begin{equation}
\langle \psi_A(t)\psi^\dagger_A(0)\rangle_{2 n} = \frac{(-1)^n}{(2 n)! \hbar^{2 n}} \sum_{ \{ \eta_{j} = \pm 1 \}} \eta_1 \ldots \eta_{2 n}
\int_ d t_1\,\ldots t_{2 n} \langle T_K \psi_A(t^-)\psi^\dagger_A(0^+)
H_{BS}(t_1^{\eta_1}) \ldots \,H_{BS}(t_{2 n}^{\eta_{2 n}})\rangle.
\end{equation}
where we dropped the $L$ dependence for easiness.
We extract the dominant behavior by having again pairs of times close to each other and inbetween $0$ and $t$. The correction takes the form
\begin{equation}
\begin{split}
\frac{1}{2 \pi v_F}  \frac{1}{a+i t}   \frac{(-1)^n r^{2 n}}{(2 n)! (2 \pi)^{2 n}} \paf{(N+1) e^\gamma E_C}{\pi \hbar}^{\frac{2 n}{N+1}}  \sum_{ \{ \eta_{j} = \pm 1 \}} \eta_1 \ldots \eta_{2 n} 
\int_{[0,t]^{2n}} d t_1\,\ldots t_{2 n} \sum_{\{\varepsilon_{j} = \pm 1 \}} e^{i (e V/\hbar) \sum_j    \varepsilon_j t_j} \\  \delta_{\varepsilon_1 + \ldots + \varepsilon_{2 n},0} 
 \left [ \sum_{\{ k, \ell \} \atop \varepsilon_{k} = 1, \varepsilon_{\ell} = -1 } \prod_{k=1}^n \prod_{\ell=1}^n \left( \frac{1}{a+i (t_{k}-t_{\ell})\,\chi_{\eta_{k}\eta_{\ell}(t_{k}-t_{\ell})}} \right)^2
\frac{[a+i (t-t_k) \eta_k  ]\,[a- i t_\ell \eta_\ell]}{[a+i (t-t_\ell) \eta_\ell ]\,[a- i t_k \eta_k]}
 \right ]^{\nu}  
\end{split}
\end{equation}
where we have used an extension of Wick's theorem for fractional vertex operators. The $\delta_{\varepsilon_1 + \ldots + \varepsilon_{2 n},0} $ term imposes that, among the set $(\varepsilon_1, \ldots, \varepsilon_{2 n})$, half of them are $+1$, the other half $-1$. Once the choice of positive and negative ones is done, the notation $\{ k, \ell \} \atop \varepsilon_{k} = 1, \varepsilon_{\ell} = -1$ includes all possible pairings that take one partner with $\varepsilon=1$ and the other with $\varepsilon=-1$. There are $n!$ such pairings.

To make further progress, we identify that the integral is dominated by approximately disjoint domains where the time pairs $(t_k,t_\ell)$ are in proximity. For each pair, we switch the integration variables to the relative time $t_k-t_\ell$ and the center of mass. All terms give the same contribution and we just multiply the result by the number of pairings $n !$ and the number of choices ${2 n}\choose{n}$ for all possible $(\varepsilon_1, \ldots, \varepsilon_{2 n})$ satisfying that their sum is zero. Integrating for each pairing over its time center of mass, we are able to factorize the result into
\begin{equation}
\frac{1}{2 \pi v_F}  \frac{1}{a+i t}   \frac{(-1)^n r^{2 n}}{(2 \pi)^{2 n}} \paf{(N+1) e^\gamma E_C}{\pi \hbar}^{\frac{2 n}{N+1}} \frac{t^n}{n!} 
\left( \sum_{\eta_1,\eta_2} \eta_1 \eta_2 \int_{-\infty}^{+\infty} d \tau  \,  \frac{e^{i e V \tau/\hbar}}{\left [ a+i \tau \chi_{\eta_{1}\eta_{2}} (\tau) \right]^{2 \nu}} e^{i \pi \nu (\eta_1-\eta_2)} \right)^n
\end{equation} 
or 
\begin{equation}
\frac{1}{2 \pi v_F}  \frac{1}{a+i t} \frac{1}{n!}  \left( - \frac{r^2}{(2 \pi)^2} \paf{(N+1) e^\gamma E_C}{\pi \hbar}^{\frac{2}{N+1}} {\cal K} (t)  \right)^n = \frac{1}{2 \pi v_F}  \frac{1}{a+i t} \frac{1}{n!} \left[ - \frac{I_A}{e^*}  \left( 1 - e^{- 2 i \pi \nu}  \right) \, t \right]^n
\end{equation} 
Summing over all orders $n$ of perturbation, we use the Taylor expansion of the exponential and obtain the final result
 \begin{equation}
\langle \psi_A(L,t)\psi^\dagger_A(L,0)\rangle= \frac{1}{2 \pi v_F}  \frac{e^{- \frac{I_A}{e^*} \left( 1 - e^{- 2 i \pi \nu}  \right)\, t }}{a+i t}.
\end{equation}

\end{document}